\documentstyle[psfig,aps,prl,multicol,eqsecnum]{revtex}
\begin{document}

\title{Entropic bounds on coding for noisy quantum channels}

\author{Nicolas J. Cerf$^{1,2}$}
\address{$^1$W. K. Kellogg Radiation Laboratory,
California Institute of Technology, Pasadena, California 91125\\
$^2$Information and Computing Technologies Research Section,
Jet Propulsion Laboratory, Pasadena, California 91109}

\date{Received 9 July 1997}

\draft
\maketitle
\begin{abstract}

In analogy with its classical counterpart, a noisy quantum channel
is characterized by a {\it loss}, a quantity that depends on the
channel input and the quantum operation performed by the channel.
The loss reflects the transmission quality: if the loss is zero,
{\it quantum} information can be perfectly transmitted 
at a rate measured by the quantum source entropy. By using {\it block
coding} based on sequences of $n$ entangled symbols, the {\it average} loss
(defined as the overall loss of the joint $n$-symbol channel divided by
$n$, when $n\to \infty$) can be made lower than the loss for a {\em single}
use of the channel. In this context, we examine several upper bounds
on the rate at which quantum information can be transmitted reliably
via a noisy channel, that is, with an asymptotically vanishing
average loss while the {\em one-symbol} loss of the channel is non-zero.
These bounds on the channel capacity rely
on the entropic Singleton bound on quantum error-correcting codes
[Phys. Rev. A {\bf 56}, 1721 (1997)].
Finally, we analyze the Singleton bounds when the noisy
quantum channel is supplemented with a classical auxiliary channel.

\end{abstract}
\pacs{PACS numbers: 03.65.Bz, 03.67.Hk, 89.70.+c
      \hfill KRL preprint MAP-215}

\begin{multicols}{2}[]
\narrowtext

\section{Introduction}

Within recent years, the quantum theory of information and
communication has undergone a dramatic evolution (see, e.g.,
\cite{bib_phystoday}). Major progress has been made toward
the extension to the quantum regime of the classical theory of information
pioneered by Shannon~\cite{bib_shannon}. In particular,
the use of {\it quantum} communication channels in order to transmit
not only classical information but also intact quantum states
(or {\it quantum} information) has received a considerable
amount of attention, following the proof of the
quantum analog of Shannon's fundamental theorem for noiseless coding 
by Schumacher~\cite{bib_noiseless}. It has been shown that
the von Neumann entropy plays the role of a quantum information-theoretic
entropy in the sense that it characterizes the minimum
amount of quantum resources (e.g., number of quantum bits) that is
necessary to code an ensemble of quantum states with an asymptotically
vanishing distortion in the absence of noise.
This result suggests that a general quantum theory of information,
paralleling Shannon theory, can be developed
based on this concept. While such a full theory does not exist as of yet,
a great deal of effort has been devoted to this issue over the last
few years, and several fundamental results have been obtained,
ranging from entanglement-based communication schemes~\cite{bib_superdense}
to quantum error-correcting codes~\cite{bib_qecodes}.
In particular, a substantial amount of work has been devoted
recently to the transmission of arbitrary states (or quantum information)
through {\it noisy} quantum channels
(see, e.g., \cite{bib_bdsw,bib_schum,bib_lloyd,bib_channel}).
A quantum state processed by such
a channel undergoes {\it decoherence} by
interacting with an external system or environment, which 
effects an alteration of quantum information. 
A natural question that arises in this context concerns the possibility
of transmitting quantum information {\it reliably}, in spite of quantum noise,
if it is suitably encoded as sequences of quantum bits
in analogy with the standard construction used for classical
channels. More specifically, a fundamental issue is to understand
the quantum analog of Shannon's noisy channel coding theorem and
to define the {\it capacity} of a noisy quantum channel, i.e., 
an upper limit to the amount of quantum information that can be
processed with an arbitrarily high fidelity. While several
attempts have been made to define a quantum analog of Shannon
mutual information that would be a natural candidate
for such a quantum measure of capacity (see the concepts of coherent 
information~\cite{bib_schum,bib_lloyd} or von Neumann mutual
entropy~\cite{bib_channel,bib_neginfo}),
the problem of characterizing in general the capacity of 
a noisy quantum channel is still unsolved.
\par

The purpose of this paper is to further clarify the description of
noisy quantum channels centered on the von Neumann mutual entropy
(see~\cite{bib_channel}). It has been shown recently that
a consistent information-theoretic framework that closely parallels Shannon's
construction can be developed, based on von Neumann 
{\it conditional} and {\it mutual} 
entropies~\cite{bib_neginfo,bib_oviedo,bib_physcomp,bib_accessible}.
The central peculiarity of this framework is that it involves 
{\it negative} conditional entropies in order to account for
quantum non-local correlations between entangled variables. This is
in contrast with Shannon information theory in which marginal and
conditional entropies are all non-negative quantities.
Negative quantum conditional entropies simply reflect the
non-monotonicity of the von Neumann entropy~\cite{bib_wehrl} (the entropy of a
composite system can be lower than that of its components 
if the latter are entangled).
The resulting information-theoretic formalism provides grounds
for the quantum extension of the usual algebraic relations between Shannon
entropies in multipartite 
systems~\cite{bib_oviedo,bib_physcomp,bib_accessible}.
Surprisingly, many concepts of Shannon theory
can be straightforwardly translated to the quantum regime
by extending the range for 
quantum (conditional and mutual) entropies with respect to the
classical one in order to encompass entanglement~\cite{bib_neginfo}.
This is very helpful in analyzing
quantum information processes in a {\em unified} framework, paralleling
Shannon theory. For example, 
entanglement-based quantum communication processes~\cite{bib_neginfo},
quantum channels~\cite{bib_channel}, and quantum error-correcting 
codes~\cite{bib_cerfcleve}
can be described along these lines.
\par

In this paper,
we focus on the application of this information-theoretic framework 
to the issue of finding upper bounds on the capacity of quantum codes
and quantum channels. In Section II, we outline the general treatment
of noisy quantum channels based on quantum entropies~\cite{bib_channel},
and extend it to the characterization of consecutive uses
of a quantum memoryless 
channel (cf. the notions of one-symbol and average loss explained in
Section  IID). This provides a simple framework to consider
{\em block coding} with quantum channels. Note that,
just as in Shannon information theory, quantum entropic considerations
alone do not result in {\it constructive} methods for building codes.
Rather, they are useful to derive bounds
on what can possibly be achieved or not, from basic principles.
Accordingly, we analyze in Section III several upper bounds 
(based on the Singleton bound on quantum codes~\cite{bib_cerfcleve})
for standard quantum channels such as the
quantum erasure or depolarizing channel. This
confirms bounds on the quantum capacity that were derived
otherwise, but places this problem in a unified context.
Finally, we examine in Section IV the extension of this quantum
entropic treatment
of noisy quantum channels to the case where an auxiliary {\em classical}
channel is available. Quantum teleportation appears then as a special case
of this construction when no block coding is applied.

\section{Entropic characterization of noisy quantum channels}
\subsection{Notations}

Let us start by summarizing the basic definitions 
that will be useful in the rest of this paper when considering
noisy quantum channels. 
The entropy of a quantum system $X$ (of arbitrary dimension)
is defined as the von Neumann entropy of the density operator $\rho_X$
that characterizes the state of $X$, i.e.,
\begin{equation}
S(X)=S[\rho_X] \equiv -{\rm Tr}(\rho_X \log_2 \rho_X)
\end{equation}
It can be viewed as the uncertainty about $X$ in the sense that
it measures (asymptotically) the minimum number of quantum bits (qubits)
necessary to specify $X$~\cite{bib_noiseless}. 
This definition can be extended to the notions of {\em conditional}
and {\em mutual} von Neumann entropies, based on a simple
parallel with their classical counterparts which is motivated 
in~\cite{bib_neginfo,bib_oviedo,bib_physcomp}. For a bipartite system
$XY$ characterized by $\rho_{XY}$, the {\em conditional} von Neumann 
entropy is
\begin{equation}
S(X|Y) = S(XY) - S(Y)
\end{equation}
while the {\em mutual} von Neumann entropy is
\begin{eqnarray}
S(X{\rm:}Y)&=& S(X)-S(X|Y)\nonumber\\ 
           &=& S(Y)-S(Y|X)\nonumber\\
           &=& S(X)+S(Y)-S(XY)
\end{eqnarray}
where $S(XY)$ is calculated from $\rho_{XY}$, while $S(X)$ and $S(Y)$
are obtained from the reduced density operators
$\rho_X={\rm Tr}_Y(\rho_{XY})$ and
$\rho_Y={\rm Tr}_X(\rho_{XY})$. Subadditivity of quantum
entropies implies $S(X{\rm:}Y)\ge 0$, where the equality holds if $X$
and $Y$ are independent (i.e., $\rho_{XY}=\rho_X\otimes \rho_Y$).
Note that, when
$S(XY)=0$ (i.e., the joint system $XY$ is in a pure state),
we have $S(X{\rm:}Y)=2S(X)=2S(Y)$ as a consequence of the Schmidt
decomposition. This property will be useful in the following.
Several quantum entropies can also be defined for characterizing
{\em multipartite} quantum systems. 
Consider, for instance, a tripartite system
$XYZ$. The von Neumann conditional mutual entropy (of $X$ and $Y$,
conditionally on $Z$) can be defined as
\begin{eqnarray}
S(X{\rm:}Y|Z)&=&S(X|Z)-S(X|YZ)\nonumber\\
             &=&S(X|Z)+S(Y|Z)-S(XY|Z) \nonumber\\
             &=&S(XZ)+S(YZ)-S(Z)-S(XYZ)
\end{eqnarray}
in perfect analogy with the classical expressions. Note that the strong
subadditivity of quantum entropies implies 
$S(X{\rm:}Y|Z)\ge 0$~\cite{bib_physcomp}. We can also define
the von Neumann {\em ternary} mutual entropy as
\begin{equation}
S(X{\rm:}Y{\rm:}Z)= S(X{\rm:}Y)-S(X{\rm:}Y|Z)
\end{equation}
Note that, if $S(XYZ)=0$ (i.e., the ternary system is in a pure
state), then $S(X{\rm:}Y{\rm:}Z)=0$~\cite{bib_physcomp},
or, equivalently, $S(X{\rm:}Y)=S(X{\rm:}Y|Z)$, 
a property which is very useful in the analysis of
quantum channels. Also, {\it chain rules} for quantum entropies
can be written, such as
\begin{equation}
S(X{\rm:}YZ)=S(X{\rm:}Y)+S(X{\rm:}Z|Y)
\end{equation}
which parallel the classical relations~\cite{bib_physcomp}.
The motivation for building such a quantum entropic framework is that 
it provides an {\em information-theoretic} formulation of quantum
entanglement in multipartite systems, 
unified with Shannon's description of classical
correlation. It is an extension of Shannon's formalism 
beyond its original range, as reflected for example by
the fact that the quantum mutual entropy can reach 
{\em twice} the maximum value allowed for classical
entropies~\cite{bib_neginfo}, that is,
\begin{equation}
0 \le S(X{\rm:}Y) \le 2 \min[S(X),S(Y)]
\end{equation}
This factor 2 appears in many quantum information-theoretic 
relations (see below), and originates from the Araki-Lieb inequality
for quantum entropies~\cite{bib_neginfo,bib_oviedo,bib_physcomp}.
\par

\subsection{Quantum mutual entropy, loss, and noise}

Let us now outline the entropic treatment of a
noisy quantum channel (see also Ref.~\cite{bib_channel}). Such a
treatment explicitly displays the correspondence with the
standard description of noisy classical channels (see Appendix A),
thereby unifying classical and quantum channels.
Our description involves three quantum systems of arbitrary dimensions:
$Q$ (the quantum system whose processing
by the channel is concerned), $R$ (a ``reference'' system
which $Q$ is initially entangled with), and $E$ (an external system 
or environment which $Q$ is interacting with in the noisy channel).
More specifically, we assume that $Q$ is initially
entangled with $R$, so that the joint state of $Q$ and $R$ 
is the {\em pure} state $|\Psi_{RQ}\rangle$. 
We may as well regard $Q$ as a quantum
source, being initially in a mixed state $\rho_Q$
(realized by a given ensemble of quantum states 
associated with some probability distribution).
The ``purification'' of $\rho_Q$ into $|\Psi_{RQ}\rangle$
can always achieved by extending the Hilbert space ${\cal H}_Q$ to 
${\cal H}_{RQ}$,
so that we have $\rho_Q={\rm Tr}_R (|\Psi_{RQ}\rangle\langle \Psi_{RQ}|)$.
The corresponding reduced von Neumann entropies are
\begin{equation}
S(R)=S(Q) \equiv S
\end{equation}
where $S$ is called the {\em source entropy}.
In the dual picture where an
{\it arbitrary} pure state of $Q$ (rather than entanglement)
is sent through the channel, $S$ then
measures the ``arbitrariness'' of $Q$ (it can be viewed as 
the average number of quantum bits that are to be processed by the channel
in order to transmit the state of $Q$).
In what follows, we prefer to
consider a quantum input $Q$ that is entangled with $R$, so that the
preservation of entanglement---rather than of arbitrary 
states---will be the central feature of a quantum transmission channel.
The initial {\em mutual entropy} to be transmitted is thus
\begin{equation}
S(R{\rm:}Q)=2S
\end{equation}
that is, {\em twice}\footnote{Note that this factor
{\em two} reflects a fundamental difference between classical and quantum
channels (see Appendix A for comparison). Such a factor is omnipresent
in the quantum information-theoretic relations between
entropies~\cite{bib_physcomp}.}
the source entropy.
\par

\begin{figure}
\caption{Schematic representation of the quantum operation effected
by a noisy quantum channel. The quantum system $Q$ is initially
entangled with the reference $R$, with a mutual entropy of twice
the source entropy $S$ (this is indicated by a dashed line). Then $Q$
decoheres by interacting with an environment $E$ (initially in a pure state
$|0\rangle$). The
entropy Venn diagram summarizes the entropic relations between
$Q'$ (output of the quantum channel),
$R'$ (reference), and $E'$ (environment) {\it after} decoherence.
The three parameters, $I$, $L$, and $N$, denote the von Neumann
mutual entropy (quantum information), the loss, and the noise, respectively. }
\vskip 0.25cm
\centerline{\psfig{figure=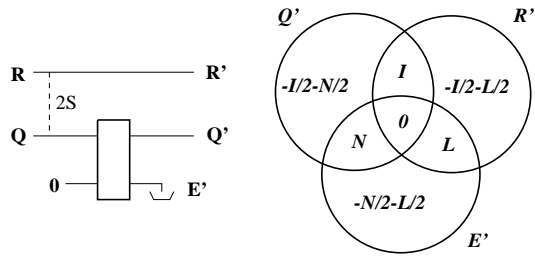,width=2.75in,angle=-90}}
\label{fig_diagram}
\vskip -0.25cm
\end{figure}

When it is processed by the channel,
$Q$ interacts with $E$ (assumed to be initially
in a pure state $|0\rangle$) according to the unitary transformation
$U_{QE}$, inducing {\it decoherence}. This describes the most general
(trace-preserving) operation of a quantum channel that is allowed by quantum
mechanics. Roughly speaking, the resulting {\it noisy} quantum channel
is such that, typically, only a fraction
of the initial entanglement with $R$ can be recovered
after having been processed by the channel
(the rest of the entanglement with $R$ is lost, in the sense that it is
transferred to the environment). More specifically,
the decohered quantum system after interaction with $E$, denoted as $Q'$,
is in the state
\begin{equation}
\rho_Q' = {\rm Tr}_E \left( U_{QE} (\rho_Q \otimes |0\rangle\langle 0|)
U_{QE}^{\dagger} \right)
\end{equation}
where $\rho_Q$
is the initial state of $Q$ (with source entropy $S$). The completely
positive linear map
$\rho_Q \to \rho_Q'$ corresponds to the ``quantum operation''
performed by the noisy channel~\cite{bib_schum}.
After such an environment-induced decoherence,
the joint system $R'Q'E'$ is in the state
$|\Psi_{R'Q'E'}\rangle=(1_R \otimes U_{QE}) |\Psi_{RQ}\rangle
|0_E\rangle$ whose entropy Venn diagram
is represented in Fig.~\ref{fig_diagram}
(the primes refer to the systems {\em after} decoherence). 
Note that, as the reference is not 
involved in decoherence, we have $R' \equiv R$.

The entropy diagram of $R'Q'E'$ depends on three
parameters, the von Neumann {\em mutual} entropy (or the quantum
information) $I$, the {\it loss} $L$, and the {\it noise} $N$,
these quantities being defined in analogy with their classical
counterparts:
\begin{eqnarray}
I&=&S(R{\rm:}Q') \label{eq_III}\\
L&=&S(R{\rm:}E'|Q')=S(R{\rm:}E')  \label{eq_LLL}\\
N&=&S(Q'{\rm:}E'|R)=S(Q'{\rm:}E') \label{eq_NNN}
\end{eqnarray}
The classical correspondence can be made fully explicit by including
an environment in the description of a classical channel,
as shown in~\cite{bib_channel}. The second equality in 
Eqs.~(\ref{eq_LLL}) and (\ref{eq_NNN}) has no classical analog,
and results from the vanishing of the ternary mutual entropy
$S(R{\rm:}Q'{\rm:}E')$ (see~\cite{bib_channel,bib_cerfcleve}).
Physically, the quantum information $I$ corresponds to the residual mutual
entropy between the decohered quantum output $Q'$ and the reference
system $R$ that purifies the quantum input $Q$. 
The loss $L$ is the mutual entropy that has arisen
between the environment after decoherence $E'$ and the reference
system $R$, while the noise $N$ is the mutual entropy between the
decohered quantum output $Q'$ and the environment $E'$.
Note that $I$, $L$, and $N$ can be written
as a function of reduced entropies only, without explicitly involving
the environment $E$ in the discussion, by making
use of the Schmidt decomposition of the state of $R'Q'E'$,
namely $S(E')=S(RQ')$:
\begin{eqnarray}
I&=& S(Q)+S(Q')-S(RQ') \\
L&=& S(Q)+S(RQ')-S(Q') \\
N&=& S(Q')+S(RQ')-S(Q)
\end{eqnarray}
It can also be shown that these three quantities are in fact
independent of the
choice of the reference system $R$ whenever the latter purifies the
quantum input $Q$, so that they provide a most concise entropic
characterization of information flow in the channel.
They depend in general on the channel input (i.e., $\rho_Q$) {\em and} 
on the quantum operation performed by the channel
(i.e., the completely positive trace-preserving map on $Q$ that is specified
by $U_{QE}$ in the joint space of $Q$ and $E$). This exactly
parallels the situation for the analog classical quantities.
The information $I$, loss $L$, and noise $N$ of
a classical channel of input $X$ and output $Y$ 
(see Appendix A) indeed depend 
on the input distribution $p(x)$ and on the channel ``operation'' 
characterized by $p(y|x)$.
\par

Among these three quantities, only $I$ and $L$ are relevant 
as far as (forward) information transmission through the channel
is concerned (the noise $N$ plays
a role in the description of the ``reverse'' channel,
just as for classical channels). Indeed,
information processing is characterized by the balance between
the von Neumann mutual entropy and the loss, these two quantities
always summing to {\em twice} the source entropy:
\begin{equation}  \label{eq_I+L}
I + L = 2 S(Q) \equiv 2S
\end{equation}
The mutual entropy $I=S(R{\rm:}Q')$ 
represents the amount of the initial mutual 
entropy with respect to $R$ (i.e., $2S$) that has been processed 
by the channel, while
the loss $L=S(R{\rm:}E')$ corresponds to the fraction of it
that is unavoidably lost in the
environment. If the channel is {\it lossless} ($L=0$), then $I=2S$,
so that the interaction 
with the environment can be perfectly ``undone'', and the initial
entanglement of $Q$ can be fully recovered 
by an appropriate decoding~\cite{bib_schum,bib_channel}.
(Equivalently, this means that an arbitrary initial state
of $Q$ can be recovered without error.) This can be understood by
noting that $R$ does not become entangled {\em directly} with the environment
in a lossless channel, but only via the output $Q'$ 
(see Fig.~\ref{fig_diagram} when $L=0$). An operation
on $Q'$ only (namely, the decoding operation) is enough to transfer the
entanglement with $E'$ (measured by the noise $N$) to an ancilla,
while preserving the entanglement $2S$ with $R$. 
\par

Thus, if $L=0$, a perfect 
transmission of information ({\it including} quantum information)
can be achieved through the channel by applying an appropriate decoding. 
When $I=0$, on the other hand,
no information at all (classical {\it or} quantum) can be processed by
the channel. This is the case, for example, of the quantum
depolarizing channel with $p=3/4$ (see Section~IIID). In between these
limiting cases, classical information (and, up to some
{\em restricted} extent, quantum information) can be reliably
transmitted at the expense of a decrease in the rate
by making use of {\em block coding}. The analysis
of such a transmission of quantum information
immune to noise is the main focus of this paper.
\par

For completeness, let us mention
that a channel with $N=0$ is the quantum analog of a {\em deterministic}
channel~\cite{bib_ash},
that is, a channel where the input fully determines the output
(see Appendix A). The quantum output $Q'$ is indeed not directly entangled
with $E'$ but only via $R$, which implies that its entanglement with $R$
remains intact (see Fig.~\ref{fig_diagram} when $N=0$).
This does not mean, however, that perfect
error correction is achievable, as an operation on the reference $R$ 
is needed to recover the initial entanglement $2S$ between $Q$ and $R$.
A channel which is both lossless ($L=0$)
and deterministic ($N=0$) is called {\it noiseless}; its action on $Q$
is the identity operator (or any {\it fixed} unitary operator).
For example, the overall channel including a noisy quantum channel
along with the encoder and decoder is obviously noiseless if 
perfect error correction is achieved. (In other words, the decoder
is used to eliminate the quantum noise $N\ne 0$ by transferring 
the entanglement with $E$ to an ancilla, which then makes the overall
channel noiseless {\em provided} that $L=0$.)
It is worth noting here that the noise $N$ and the loss $L$ play symmetric
roles when considering the ``reverse'' channel obtained by interchanging
the input and output. (This is true for classical channels as well.) 
More specifically, $N$ and $I$ always sum
to twice the {\em output} entropy,
\begin{equation}
I+N= 2 S(Q')
\end{equation}
in analogy with Eq.~(\ref{eq_I+L}). Roughly speaking,
$N$ plays the role of the loss of the reverse channel,
as shown in Sec.~IIC.

\subsection{Properties of quantum $I$, $L$, and $N$}

The above entropies for a noisy quantum channel can be
shown to fulfill several properties, akin to classical ones, which make
them reasonable quantum measures of information, loss, or noise 
(see also Ref.~\cite{bib_channel}). First, the quantum mutual entropy
$I$ can be shown to be {\em concave} in the input $\rho_Q$ for a fixed
channel, i.e., a fixed quantum operation $\rho_Q\to\rho_{Q'}$ or a
fixed $U_{QE}$. Therefore, any local maximum of $I$ is the absolute
maximum, that is, the von Neumann capacity of the channel.
This parallels the concavity of the Shannon mutual entropy
$H(X{\rm:}Y)$ in the input probability distribution $p(x)$ for a fixed
channel, i.e., fixed $p(y|x)$~\cite{bib_cover}.
Second, $I$ is {\em convex} in the output
$\rho_{Q'}$ for a fixed input $\rho_Q$. This property will be used in
the next Section when considering a ``probabilistic'' channel (the
effective channel resulting from the probabilistic use of a family of 
channels). It is the quantum analog
of the property that the information $H(X{\rm:}Y)$ processed
by a classical channel is a convex
function of $p(y|x)$ for a fixed $p(x)$~\cite{bib_cover}.
These two properties are simple to prove by reexpressing the
von Neumann mutual entropy $I$ as
\begin{eqnarray}
S(R{\rm:}Q') 
&=& S(Q'E') + S(Q') - S(E') \nonumber\\
&=& S(Q') + S(Q'|E') \label{eq_proofconcavity}
\end{eqnarray}
or as
\begin{eqnarray}
S(R{\rm:}Q')
&=&  S(R) + S(Q') - S(RQ') \nonumber\\
&=& S(R)-S(R|Q')      \label{eq_proofconvexity}
\end{eqnarray}
If the input $\rho_Q$ is a convex combination of density operators
while the channel is fixed, it is easy to see that $\rho_{QE}$ and therefore
$\rho_{Q'E'}$ are also convex combinations (as the channel operation
is linear). 
Since the conditional entropy $S(Q'|E')$ is concave in a convex combination
of $\rho_{Q'E'}$ while $S(Q')$ is concave in $\rho_{Q'}$~\cite{bib_wehrl},
Eq.~(\ref{eq_proofconcavity})
implies the {\em concavity} of the quantum mutual entropy $I$
in the input for a fixed channel.
The second property can be proven the same way by noting that,
if we have a ``probabilistic'' channel---a convex combination
of quantum channels---acting on a fixed input,
then $\rho_{RQ'}$ is a convex combination of density operators
while $\rho_R$ is constant.
Thus, Eq.~(\ref{eq_proofconvexity}) together with
the concavity of the conditional entropy $S(R|Q')$ 
in a convex combination of $\rho_{RQ'}$ implies that 
the quantum mutual entropy $I$ is {\em convex}
in the output for a fixed input.
\par

\begin{figure}
\caption{Schematic view of the chaining of two noisy quantum channels.
In each of them, the input state decoheres by interacting with
a (separate) environment. 
The input of the first channel is initially entangled with $R$, 
with a source entropy of $S$ (see the dashed line). The output of this
channel $Q'$ is then used as an input for the second channel.
Since $Q'$ is purified by $RE'$ (not by $R$ alone), the ``reference''
system that must be considered in the entropic characterization 
of the second channel is $RE'$. }
\vskip 0.50cm
\centerline{\psfig{figure=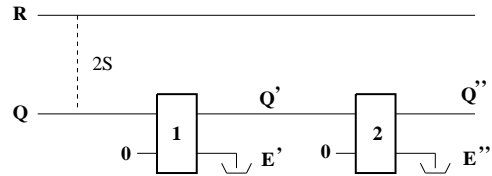,width=2.50in,angle=-90}}
\label{fig_dataproc}
\vskip -0.25cm
\end{figure}

A third important property is that the mutual entropy $I$ and
the quantum loss $L$ are {\em subadditive} when considering a channel
made of several independent quantum channels used {\em in parallel}. 
This will be shown when analyzing quantum block coding (cf. Sect. IID).
Finally, it can be proved that
$I$ obeys (forward and reverse) data-processing inequalities when
considering chained quantum channels. If we chain two channels by
using the output of the first as an input for the second
(see Fig.~\ref{fig_dataproc}), the total (1+2) channel 
$\rho_Q\to\rho_{Q'}\to \rho_{Q''}$ is characterized by
\begin{eqnarray}
I_{12}&=&S(R{\rm:}Q'') \\
L_{12}&=&S(R{\rm:}E' E'') \\
N_{12}&=&S(Q''{\rm:}E'E'')
\end{eqnarray} 
since we can regard the two environments $E'$ and $E''$ as a global
environment for this total channel .

Using the chain rule for quantum mutual entropies
$S(R{\rm:}E'E'')=S(R{\rm:}E')+S(R{\rm:}E''|E')$, and remembering that
$S(R{\rm:}E''|E')\ge 0$ as a result of strong subadditivity, we obtain
\begin{equation}  \label{eq_lossprocess}
0\le L_1 \le L_{12}
\end{equation}
where $L_1=S(R{\rm:}E')$ is the loss of the first channel 
while $L_{12}$ is the loss of the total channel. Thus, the loss
can only increase by further processing of quantum information
in the second channel.
Since $I_1+L_1=2S(Q)$ and $I_{12}+L_{12}=2S(Q)$, we obtain
the {\em forward} data-processing inequality
\begin{equation}
I_{12}\le I_1 \le 2S(Q)
\end{equation}
implying that the mutual entropy of the total channel cannot
exceed the one of the first channel. This is the quantum analog
of $H(X{\rm:}Z)\le H(X{\rm:}Y)\le H(X)$
for chained classical channels $X\to Y\to Z$~\cite{bib_cover}.
\par

Now, if we use the chain rule 
$S(Q''{\rm:}E'E'')=S(Q''{\rm:}E'')+S(Q''{\rm:}E'|E'')$ 
together with strong subadditivity, we obtain
\begin{equation}  \label{eq_noiseprocess}
0\le N_2 \le N_{12}
\end{equation}
where $N_2=S(Q''{\rm:}E'')$ is the noise of the second 
channel while $N_{12}$ is the noise of the total channel.
As $I_2+N_2=2S(Q'')$ and $I_{12}+N_{12}=2S(Q'')$, we obtain
the {\em reverse} data-processing inequality
\begin{equation}
I_{12}\le I_2 \le 2 S(Q'')
\end{equation}
where $I_2=S(RE'{\rm:}Q'')$ is the mutual entropy processed
by the second channel. (Note that the ``reference'' system that
purifies the input $Q'$ of the second channel is $RE'$.) This parallels the
classical inequality $H(X{\rm:}Z)\le H(Y{\rm:}Z)\le H(Z)$
for chained channels~\cite{bib_cover}.
Eqs.~(\ref{eq_lossprocess}) and (\ref{eq_noiseprocess})
emphasize that the loss $L$ and the noise $N$
play a symmetric role in this entropic description if one interchanges 
the input and the output of the quantum channel (``time-reversal''),
just as for classical channels. This is
reflected by the symmetry between the forward and the reverse 
data-processing inequalities.

\subsection{One-symbol loss and average loss}

The central idea of classical error correction by block coding is to
introduce {\it correlations} between the bits that make a block,
in order to have redundancy in the transmitted flow of data. This can make
the transmission asymptotically immune to errors,
up to some level of noise.
In quantum error-correcting codes, the qubits that form a block are
{\it entangled} in a specific way, so that a partial alteration
due to decoherence can be recovered~\cite{bib_qecodes}. 
Even though entanglement gives rise
to some qualitatively new features (see~\cite{bib_cerfcleve} for
a detailed analysis), the objective is similar. Namely,
when block coding is used, i.e., when say $k$ ``logical'' qubits are encoded
into blocks of $n$ ``physical'' qubits, it is possible to
achieve a situation where the {\it overall} loss of the joint ($n$-bit)
channel is arbitrarily small, while the loss for individual qubits
(for each use of the channel) is finite. 
In analogy with the classical construction, if 
blocks of $n$ qubits that are initially entangled with respect to $R$ (with a
mutual entropy $2k$) can be transmitted through the channel
with an asymptotically vanishing overall loss, we say that
the channel processes $2k/n$ bits of entanglement {\it per qubit}.
Equivalently, the channel is transmitting at a {\it rate} $R=k/n$
(on average, $k$ arbitrary binary quantum states can be transmitted for $n$
transmitted qubits). The maximum rate at which quantum information
can be reliably sent through the noisy channel is defined as
the quantum {\em channel capacity}. 
(This maximum has to be taken over all possible
coding schemes, and for $n\to \infty$.) 
Whether a good (and operational) definition
of such a ``purely quantum'' channel capacity exists is currently an open
question. In the following, we restrict ourselves 
to the issue of finding upper
bounds on the rate of perfect quantum information transmission
(and therefore on such a ``purely quantum'' capacity). 
\par

\begin{figure}
\caption{Schematic view of a memoryless quantum channel. This channel is
used $n$ times, but the environment is ``reset'' after each use. This
can be viewed as $n$ parallel (independent) channels, each one being used for
one of the input symbols. The $n$ input symbols ($Q_1,Q_2,\cdots Q_n$)
are initially entangled with $R$ (as indicated
by a dashed line), with a joint source entropy of $S$.}
\centerline{\psfig{figure=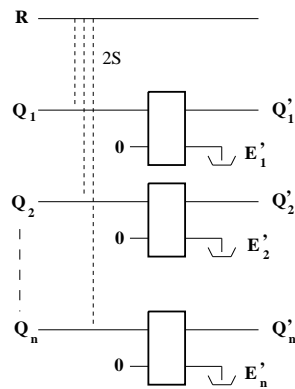,width=1.50in,angle=-90}}
\label{fig_nchannels}
\vskip -0.25cm
\end{figure}

Let us consider the asymptotic use of a quantum discrete memoryless channel,
where $n$ (tending to infinity) qubits are transmitted 
sequentially.\footnote{Throughout this paper, we use indistinctly
the terms qubit or symbol to denote the quantum state that is sent
in a single use of the channel. As a matter of fact, the reasoning is
totally general, and applies to quantum states (or symbols) in 
a Hilbert space of arbitrary dimension $\ge 2$.} Each qubit
may decohere due to an environment (quantum noise), the exact interaction
depending on the considered noise model. The important
point is that the environment for each qubit is initially {\it independent} 
of the one interacting with every other qubit. Thus the information process
can be viewed as $n$ sequential uses of a quantum {\it memoryless} channel
(the environment being ``reset'' after each use) or, equivalently,
as $n$ {\em parallel} independent channels processing one qubit each
(see Fig.~\ref{fig_nchannels}).
We assume that the set of $n$ input symbols ($Q_1$, \dots $Q_n$) are
initially entangled with $R$, so that
$S(R{\rm:}Q_1 \cdots Q_n)=2S$ and $S(R)=S(Q_1 \cdots Q_n)=S$.
If we consider these $n$ symbols as
the single input of a joint $n$-bit channel
$Q_1\cdots Q_n \to Q_1'\cdots Q_n'$, information transmission
is described by the mutual entropy
\begin{eqnarray}
I&=&S(R{\rm:}Q_1' \cdots Q_n')  \nonumber \\
 &=&S(Q_1 \cdots Q_n)+S(Q_1' \cdots Q_n')-S(E_1' \cdots E_n')\nonumber\\
\label{eq_Itot}
 &=&S(Q_1' \cdots Q_n'|E_1' \cdots E_n') + S(Q_1' \cdots Q_n')
\end{eqnarray}
and the loss
\begin{eqnarray}
L&=&S(R{\rm:}E_1' \cdots E_n')  \nonumber \\   
 &=&S(Q_1 \cdots Q_n)+S(E_1' \cdots E_n')-S(Q_1' \cdots Q_n')\nonumber\\
\label{eq_Ltot}
 &=&S(E_1' \cdots E_n'|Q_1' \cdots Q_n') + S(E_1' \cdots E_n')
\end{eqnarray}
where we have made use of the conservation of entropy imposed
by the unitarity of the global interaction with the $n$ environments
$E_1$, \dots $E_n$.
Obviously, we have $I+L=2S(Q_1 \cdots Q_n)=2S(R)$, which is twice
the source entropy $S$ of the joint channel.
\par

Each individual channel $Q_i \to Q_i'$ can be described in the same
way, noting that each $Q_i$ interacts with
an environment $E_i$ (initially in a pure state $|0\rangle$)
which results in $Q_i'$ and $E_i'$. The only difference here is that
$R$ has to be supplemented with all the input $Q$'s
{\em except} $Q_i$ in order to purify $Q_i$. Thus, the mutual
entropy characterizing the $i$-th channel is
\begin{equation}
I_i=S(RQ_1\cdots Q_{i-1}Q_{i+1}\cdots Q_n {\rm:} Q_i')
\end{equation}
while the corresponding loss is
\begin{equation}
L_i=S(RQ_1\cdots Q_{i-1}Q_{i+1}\cdots Q_n {\rm:} E_i')
\end{equation}
These quantities can be re-expressed by using the fact that
the environments $E_1$, \dots $E_n$ are initially in a product state
and that the $Q_i$ and $E_i$ interact pairwise.
We have
\begin{eqnarray}
I_i &=& S(Q_i)+S(Q_i')-S(E_i') \nonumber\\
&=& S(Q_i' | E_i') + S(Q_i')   \label{eq_Ii}
\end{eqnarray}
and
\begin{eqnarray}
L_i &=& S(Q_i)+S(E_i')-S(Q_i') \nonumber\\
&=& S(E_i' | Q_i') + S(E_i')   \label{eq_Li}
\end{eqnarray}
For each channel, the loss and the mutual entropy sum to twice
the source entropy of the channel (tracing over all the other
channels): $I_i+L_i = 2 S(Q_i)$.
The subadditivity of von Neumann entropies,
\begin{equation}   \label{eq_subad1}
S(E_1'\cdots E_n') \le S(E_1') + \cdots + S(E_n')
\end{equation}
and the subadditivity of von Neumann conditional entropies
\begin{equation}   \label{eq_subad2}
S(E_1'\cdots E_n'|Q_1'\cdots Q_n') \le S(E_1'|Q_1') + \cdots + S(E_n'|Q_n')
\end{equation}
combined with Eqs.~(\ref{eq_Ltot}) and (\ref{eq_Li})
imply that the loss is {\em subadditive}:
\begin{equation}
L \le L_1 + \cdots + L_n
\end{equation}
The same reasoning can be made using Eqs.~(\ref{eq_Itot}) and (\ref{eq_Ii}) 
and interchanging the $Q_i'$'s and
$E_i'$'s in Eqs.~(\ref{eq_subad1}) and (\ref{eq_subad2}),
which results in the equivalent expression for mutual entropies:
\begin{equation}
I \le I_1 + \cdots + I_n
\end{equation}
The latter inequality corresponds to the {\em subadditivity}
of the von Neumann mutual entropy for parallel channels.
Finally, using the relation between the loss and the mutual entropy
for individual channels and for the joint channel, we obtain
\begin{equation}   \label{eq_range}
L_1+\cdots +L_n - 2M \le L \le L_1 +\cdots + L_n
\end{equation}
with $M=S(Q_1)+\cdots +S(Q_n)-S(Q_1 \cdots Q_n)\ge 0$.
Equivalently, if we define the
{\it average} loss\footnote{The average loss $l$ reflects the
effective loss {\it per qubit} processed in the noisy channel,
that is, the loss affecting the overall process 
(encoding + joint channel viewed as $n$ parallel one-bit noisy channels)
divided by the number of physical qubits $n$ when $n\to\infty$.}
of the joint $n$-bit channel as $l=L/n$ for $n\to\infty$,
we see that
\begin{equation}   \label{eq_range2}
l_1 - 2m \le l \le l_1
\end{equation}
where $m=M/n$ and $l_1=(L_1 + \cdots + L_n)/n$ is the {\em one-symbol}
loss, i.e., the loss for a single use of an individual channel averaged over 
all 1-bit channels. Thus, Eqs.~(\ref{eq_range}) or (\ref{eq_range2})
imply that the loss cannot increase by using block coding
(using parallel channels). It typically decreases by an amount
which is bounded by $2M$ (or $2m$), a quantity related to the entanglement
between the input symbols. (Note that $M=0$ if the input symbols
are independent.)
The analog construction for a classical channel is presented in
Appendix A in order to clarify the straightforward classical to quantum
correspondence.
As an example, let us consider the use of blocks of two qubits.
Assume also that the two 1-bit channels are identical, i.e.,
$L_1=L_2\equiv l_1$. The {\it average} loss
of the joint 2-bit channel, $l=L/2$, can be bounded by
\begin{equation}
l_1 - S(Q_1{\rm:}Q_2) \le l \le l_1
\end{equation}
This explicitly shows that block coding
can decrease the average loss only when the symbols are
entangled [i.e., $S(Q_1{\rm:}Q_2)>0$].
\par

Equations~(\ref{eq_range}) or (\ref{eq_range2}) allow us to derive
a simple upper bound on the maximum achievable rate by
block coding, as a function of the one-symbol loss (or mutual entropy)
for a single use of the channel. Indeed, only if the lower bound
on $L$ (or $l$) extends to zero (that is, if $2M\ge L_1+\cdots +L_n$)
is it possible that block coding
makes the joint channel perfectly immune to noise while each
1-bit channel has a non-vanishing loss. Thus, we have the necessary
condition for having a vanishing average loss ($l=0$):
\begin{eqnarray}
\lefteqn{2 S(Q_1\cdots Q_n) \le 2 S(Q_1) + \cdots + 2S(Q_n)}
\hspace{3cm} \nonumber\\
&& - L_1 - \cdots - L_n
\end{eqnarray}
As a consequence, the rate of quantum information transmission through
the joint channel, $R=S(Q_1 \cdots Q_n)/n$, is bounded
from above by {\em half} the averaged {\em one-symbol} mutual entropy
for individual channels:
\begin{equation}
R \le {I_1 + \cdots + I_n \over 2n }
\end{equation}
Thus, the (averaged) mutual von Neumann entropy characterizing
each use of the channel provides an upper bound on the achievable
rate of transmission by block coding through the noisy channel. Except for the
factor 1/2, this inequality parallels the one for a classical channel
(see Appendix A). Remember that the quantum capacity of a channel
is defined as the maximum rate that can be achieved through the
channel (over all possible input and coding schemes) with a fidelity
arbitrarily close to one. The classical analogy suggests then that
the (maximum) one-symbol von Neumann mutual entropy yields the quantum
capacity. However, this upper bound appears not to be attainable
in general (see, e.g., \cite{bib_bdsw}),
in contrast with the equivalent classical bound.
(The physical meaning of the
von Neumann mutual entropy is better understood in the context of
noisy superdense coding, as shown in \cite{bib_channel}.)
Therefore, it is necessary to derive more constraining entropic
upper bounds on $R$, which is the main concern of the rest of this paper.
\par

In the next Section, we build on the
entropic derivation of the Singleton bound for quantum codes presented
in Ref.~\cite{bib_cerfcleve}, and extend it to the treatment of
noisy quantum channels in order to find better upper
bounds on the rate of perfect quantum information transmission
(and therefore on the ``purely quantum'' capacity). 
The bounds that we derive can be attained in
some cases (e.g., for the quantum erasure channel), or not in other cases
(e.g., for the quantum depolarizing channel). It is unknown whether such
a purely entropic approach unifying classical and quantum channels
can possibly yield the best (asymptotically
attainable) upper bound, just as it is the case for classical
channels, but this is not out of the question.
This will be further investigated in future work.

\section{Entropic bounds on codes and channel capacities}

In this Section, we derive several bounds, either on quantum codes
or on quantum channel capacities, using an entropic approach
based on the Singleton bound (see Ref.~\cite{bib_cerfcleve}).

\subsection{Quantum channel subject to a $p$-bounded fraction of erasures}

We say that a quantum channel is subjected to a $p$-bounded fraction 
of erasures~\cite{bib_cleve} if,
among $n$ uses of the channel, a fraction of $pn$ qubits\footnote{When
$n\to \infty$, the number of erasures $pn$ 
can be considered as an integer without loss of generality.} (at most)
are {\em erased} (or replaced by a distinguishable third state, {\it
e.g.}, $|2\rangle$). When considering {\em erasures} (rather than errors),
the important point is that it is possible to
perform an incomplete measurement of each qubit at the output of the
channel, to check whether
it is in the $|2\rangle$ (erasure) state, or in the subspace spanned
by $|0\rangle$ and $|1\rangle$, without destroying superpositions
in the latter subspace~\cite{bib_qec}. In this error model, transmission
through the channel is considered successful if an arbitrary initial quantum
state can be perfectly recovered (or the entanglement with $R$ can be
maintained), which can obviously be achieved if one uses
a quantum $e$-erasure correcting code with $e=pn$, that is,
a code that allows any pattern of $e$ qubits of each codeword to be
erased.
The {\em rate} (i.e., the average number of logical qubits transmitted with 
arbitrarily high fidelity per physical qubit) 
of a channel subjected to a $p$-bounded fraction of error
is thus equivalent to the rate of a $((n,k))$ quantum code
correcting $e=pn$ erasures.
The rate of a $((n,k))$ code, i.e., a code mapping 
$k$ logical qubits into codewords of $n$ qubits,
is defined as $R=k/n$.
Consequently, an upper bound on the rate of quantum codes is simply
equivalent to an upper bound on the rate of a channel with this 
particular error model (or an upper bound on the capacity, 
which is the highest achievable rate through the channel). When
considering a channel, $k$ is simply the source entropy $S$ of the
joint channel (i.e., the number of arbitrary qubits that are sent).
\par

It is known that an upper bound on the Hamming distance of
{\it nondegenerate} quantum codes with fixed $n$ and $k$
can be derived from ``sphere-packing''
considerations~\cite{bib_ekert}. However,
as a bound on the rate (or capacity) of a quantum channel involves 
a maximization over all coding schemes, including those based on
degenerate codes (which have been shown to exceed the 
Hamming bound~\cite{bib_shorsmolin}), only the bounds which are
valid for all quantum codes are applicable to channels.
As proven in a previous paper~\cite{bib_cerfcleve}, an upper bound
on the rate of (nondegenerate {\it and} degenerate)
quantum codes can be derived using entropic considerations
only. This is the quantum Singleton bound: $k \le n - 2e$ (see also
\cite{bib_knilaf}). Translated
in the channel language, this implies that an upper bound on the rate
(and therefore the capacity) of a lossless ($L=0$) channel
subjected to a $p$-fraction of erasures is
\begin{equation}  \label{eq_rate_p_erasures}
R \le 1 - 2 p
\end{equation}
For completeness, we summarize the proof of the Singleton bound given
in~\cite{bib_cerfcleve}. The basic idea of the proof will be useful
in the following, when considering other channels.

\begin{figure}
\caption{Schematic representation of two possible partitions of $Q$ into
an erased piece $Q_e$ (or $Q_e'$) and unerased piece $Q_u$ (or
$Q_u'$). The ``overlap''
between the unerased pieces in both partitions is denoted by $Q^*$.
The entropic erasure-correction condition for the first partition is 
$S(R{\rm:}Q_e)=0$, while the condition
for the second one is $S(R{\rm:}Q_e')=0$.}
\vskip 0.25cm
\centerline{\psfig{figure=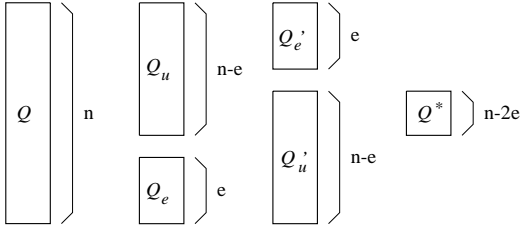,width=2.75in,angle=-90}}
\label{fig_cerfcleve}
\vskip -0.25cm
\end{figure}

As pictured in Fig.~\ref{fig_cerfcleve}, 
for each pattern of erased qubits $Q_e$, the entropic condition
$S(R{\rm:}Q_e)=0$ must be fulfilled,
so that the unerased qubits $Q_u$ emerge from a lossless channel.
This implies that the full entanglement of the codeword $Q$ (with
respect to $R$) must be ``concentrated''
in the unerased qubits $Q_u$:
\begin{equation}  \label{eq_concentr}
S(R{\rm:}Q_u)=S(R{\rm:}Q)=2S(R)
\end{equation}
Using the fact that the joint state of $RQ_uQ_e$ is pure, along with
the subadditivity of quantum entropies, we have
\begin{eqnarray}   \label{eq_SRQu}
S(R{\rm:}Q_u)&=&S(R)+S(Q_u)-S(RQ_u) \nonumber \\
             &=&S(R)+S(Q_e' Q^*) - S(Q_e) \nonumber \\
             &\le&S(R)+S(Q_e')+S(Q^*)-S(Q_e)
\end{eqnarray}
where we have divided the unerased qubits $Q_u$ into another pattern of $e$
qubits, $Q_e'$, and the remaining piece of $n-2e$ qubits, $Q^*$.
Eqs. (\ref{eq_concentr}) and (\ref{eq_SRQu}) provide the inequality
\begin{equation}
S(R)-S(Q^*) \le S(Q_e')-S(Q_e)
\end{equation}
Since this reasoning is symmetric in $Q_u$ vs. $Q_u'$, the inequality
corresponding to the division of $Q_u'$ into $Q_e$ and $Q^*$,
\begin{equation}
S(R)-S(Q^*) \le S(Q_e)-S(Q_e')
\end{equation}
must also be satisfied. Combining these two inequalities
yields the condition
\begin{equation}
S(R) \le S(Q^*) \le n-2e
\end{equation}
where the upper bound on $S(Q^*)$ simply results from the dimension of the
Hilbert space of $Q^*$. Since the encoding of a $k$-qubit arbitrary
state requires that $S(R)=S(Q)=k$, the above condition implies the
quantum Singleton bound
\begin{equation}
k \le n-2e
\end{equation}
The physical meaning of the Singleton bound is that, in order to
have $Q_e$ independent of $R$ and, at the same time, 
$Q_u' \equiv Q_e Q^*$ fully entangled
with $R$, a minimum Hilbert space for $Q^*$ (minimum
number of qubits) is necessary in order to accommodate the source entropy 
$k$.

\subsection{Quantum channel subject to a $p$-bounded fraction of errors}

Another possible error model for a quantum channel is the case
where a fraction of $p$ qubits (at most) are altered by interacting
with an environment. The difference with the previous error model
is that the location of the errors is unknown (by contrast with
erasures), i.e., there is no ``flag'' indicating which are the qubits
that have been altered. 
\par

Since a channel where the fraction of errors is bounded can be
made lossless ($L=0$) using an {\it error}-correcting code (just as for
erasures in the previous Section), it is enough to use the
correspondence between
error-correcting codes and erasure-correcting codes to derive
an upper bound on the rate of perfect transmission of quantum
information in such a channel.
In analogy with the classical situation, one can show that any
code that corrects $t$ errors is also able to correct up to $e=2t$ 
erasures~\cite{bib_grassl}. This enables us to reuse the result of the
previous Section simply by replacing $p$ by $2p$. Thus, we obtain
the Singleton upper bound on the rate (or capacity) of
a lossless ($L=0$) channel with a fixed fraction $p$ of errors:
\begin{equation}  \label{eq_rate_p_errors}
R \le 1 - 4 p 
\end{equation}
(This bound, or rather the fact that the rate of a code
is vanishing at $p=1/4$,
is originally due to Knill and Laflamme~\cite{bib_knilaf}).
To our knowledge, the only stronger bounds on $R$ for quantum codes
that have been displayed for some special cases are:
\begin{itemize}
\item $R\le H\left(1/2 +\sqrt{2p(1-2p)}\right)$ for additive (or stabilizer) 
codes, where $H$ stands for the dyadic Shannon
entropy~\cite{bib_cleve}.
This bound is based on an upper bound on classical
linear codes.\footnote{Some slightly stronger bounds have
been recently obtained in Ref.~\cite{bib_ashikhmin}.}
\item $R=0$ for $p\ge 1/6$ for general nondegenerate codes~\cite{bib_rains}.
\end{itemize}
It is worthwhile looking for improvement of
Eq.~(\ref{eq_rate_p_errors}) using an entropic approach as presented
above.

\subsection{Quantum erasure channel}

We now consider a quantum erasure channel with erasure probability $p$
(see, e.g., \cite{bib_qec}). In such a channel, each transmitted
qubit has a probability $p$ of being {\em erased} (and detectable at
the output as an ``erased'' qubit). We are interested in the
maximum rate of quantum information transmission that can be achieved
by this channel. More precisely, our aim is to derive an upper bound
on this rate using an entropic approach. It will appear
that such an entropic bound exactly coincides with
the capacity of a quantum erasure channel
recently displayed in Ref.~\cite{bib_qec}.
\par

The central point of the reasoning is to describe the joint
``probabilistic'' channel (with $n$ qubits at input and output)
as a superposition of ``binomial'' $n$-bit channels (defined below).
This allows us to make use of the convexity of the von Neumann mutual entropy 
in the output of a channel for a fixed input (see Sect. IIC)
in order to derive an upper bound on $R$. 
We consider a family of ``binomial'' $n$-bit channels $C$ (labeled by
the index $c$), each characterized by a pattern of $e$ erased qubits,
$Q_e(c)$, and the complementary pattern of $n-e$ unerased qubits, $Q_u(c)$.
The probabilistic channel of interest here
corresponds to a probabilistic use of these
channels with a {\em binomial} distribution. More precisely,
each channel $c$ with $e$ erased qubits is associated with 
a probability (or weight) $w_c=p^e (1-p)^{n-e}$, and there are obviously 
${n \choose e}$ distinct channels with $e$ erased qubits. The
superposition means, physically, that the resulting probabilistic 
joint channel
consists in using one of these $2^n$ distinct ``binomial'' $n$-bit
channels with the appropriate probability.
Thus, for a given $n$-bit input of the channel, say $\rho$, the output
can thus be written as a convex combination
\begin{equation}  \label{eq_convex}
\rho' = \sum_c w_c \rho'_c  \qquad {\rm with~}\sum_c w_c =1
\end{equation}
where $w_c$ is the weight of the $c$-th (binomial)
channel in the superposition, and
$\rho'_c$ is the output for that channel $c$.
The convexity of the von Neumann mutual entropy in the output (for a
given input) for the overall probabilistic channel implies that
\begin{equation}  \label{eq_convexity}
\underbrace{S(R{\rm:}Q')}_I \le \sum_c w_c \underbrace{S(R{\rm:}Q'_c)}_{I_c}
\end{equation}
where $I=S(R{\rm:}Q')$ is the quantum mutual entropy between the
reference $R$ and the output $Q'$ of the {\it joint} ($n$-bit)
quantum erasure channel, while 
$I_c=S(R{\rm:}Q'_c)$ stands for the mutual entropy of the output
of the $c$-th channel, $Q'_c$, with respect to $R$.
Note that, we have $S(R{\rm:}Q'_c)=S(R{\rm:}Q_u(c))$
since only the unerased qubits of the $c$-th channel contribute to mutual
entanglement with $R$ (the erased qubits are independent of $R$).
As explained in Sect. IIC, the mutual von Neumann entropy 
$S(R{\rm:}Q')=S(R)-S(R|Q')$ is convex in the output
(for a fixed input) because the conditional entropy $S(R|Q')$ is concave
in a convex combination of $\rho_{RQ'}$ [i.e., a convex combination
of quantum channels acting on a fixed input, as shown in
Eq.~(\ref{eq_convex})].
\par

It is convenient to group the
channels $C$ into several classes, according to the number of erased
qubits, $e$. 
Using Eq.~(\ref{eq_convexity}),
we can then write an upper bound on the processed quantum mutual
information in the joint ($n$-bit) erasure channel of probability $p$ as
\begin{eqnarray}\label {eq_Ip}
I(p) \le \sum_{e=0}^n (1-p)^{n-e} p^e
\sum_c S(R{\rm:}Q_u(c))
\end{eqnarray}
where the sum over $c$ spans the ${n \choose e}$ channels 
where $e$ qubits are erased.
Before deriving a simpler expression of this upper bound using
the Singleton bound, let us show that a simple relation between
$I(p)$ and $I(1-p)$ can be obtained from Eq.~(\ref{eq_Ip}).
First, note that
\begin{eqnarray}  \label{eq_several}
S(R{\rm:}Q'_c)&=&S(R{\rm:}Q_u(c)) \nonumber\\
              &=&S(R)+S(Q_u(c))-S(Q_e(c)) \nonumber\\
              &=&2 S(R) - S(R{\rm:}Q_e(c))
\end{eqnarray}
where we have used the fact that $RQ_uQ_e$ is in a
pure state for each channel $C$. Eq.~(\ref{eq_Ip}) can then be rewritten as
\begin{eqnarray}\label {eq_Ip2}
I(p) \le 2 S(R) - \sum_{e=0}^n (1-p)^{n-e} p^e
\sum_c S(R{\rm:}Q_e(c))
\end{eqnarray}
The second term in the right-hand side of Eq.~(\ref{eq_Ip2}) can be interpreted
as an upper bound on $I(1-p)$, i.e., the information processed through
a ``dual'' erasure channel of probability $1-p$ where the erased qubits are
replaced by unerased qubits and conversely. As a consequence, 
remembering that $S(R)=k$ if an arbitrary $k$-bit quantum state is sent
in the channel (i.e., the source entropy of the $n$-bit channel is $k$
bits), we have
\begin{equation}
I(p)+I(1-p) \le 2k
\end{equation}
for all $n$. Using $I(p)+L(p)=2k$,
the corresponding relation for the quantum losses of the dual 
(with probability $p$ and $1-p$) erasure channels is
\begin{equation}  \label{eq_Lp}
L(p)+L(1-p) \ge 2k
\end{equation}
(Remember that $0 \le L(p) \le 2k$.)
This implies that a {\em necessary} condition for having a perfect channel
at probability $p$, i.e., $L(p)\equiv0$, is that $L(1-p)=2k$, i.e., that the
dual channel at probability $1-p$ is ``fully erasing'' 
(no information at all---either classical or quantum---is processed
through it). Another way of expressing
this condition is by writing a lower bound on the loss of the erasing
channel
\begin{equation} \label{eq_Lp2}
L(p) \ge I(1-p)
\end{equation}
Only if no information is transmitted through 
the erasure channel of probability $1-p$, or if $I(1-p)=0$, is it possible 
that the loss of the erasure channel of probability $p$ vanishes.
This is obviously compatible with $L(0)=I(1)=0$.
Eq.~(\ref{eq_Lp}) also implies that $L(1/2) \ge k$, so that the quantum
erasure channel with probability 1/2 cannot be lossless
for a non-zero source entropy. (The fact
that it actually has a vanishing capacity---or a maximum loss---will
be shown below. This
result can also be derived from an argument based on the impossibility of
cloning, as shown in Ref.~\cite{bib_qec}.)
\par

Let us now derive a general expression for an upper bound on the
mutual entropy of the $n$-bit channel (or, equivalently, a
lower bound on the overall loss). Using Eqs.~(\ref{eq_Ip})
and (\ref{eq_several}), we have
\begin{eqnarray}\label {eq_Ip3}
\lefteqn{I(p) \le S(R) + \sum_{e=0}^n (1-p)^{n-e} p^e }
   \hspace{1cm} \nonumber \\
&\times& \sum_c \Big[ S(Q_u(c))-S(Q_e(c)) \Big]
\end{eqnarray}
We first rewrite Eq.~(\ref{eq_Ip3}) as a summation up to $n/2$ (we
assume here that $n$ is even), by combining
each channel $c$ in this sum with its dual channel
where erased qubits are unerased and conversely:
\begin{eqnarray}\label {eq_Ip4}
\lefteqn{I(p) \le S(R) + \sum_{e=0}^{n/2} \Big[ (1-p)^{n-e} p^e -
   (1-p)^e p^{n-e} \Big] }
   \hspace{1cm} \nonumber \\
&\times& \sum_c \Big[ S(Q_u(c))-S(Q_e(c)) \Big]
\end{eqnarray}
(Note that the term with $e=n/2$ is vanishing.) We now follow the
reasoning that we used earlier to derive the Singleton bound,
and group the channels in pairs ($c$ and $c'$)
which ``overlap'' in $n-2e$ qubits
denoted by $Q^*$ (see Fig.~\ref{fig_cerfcleve}). 
Thus, for a given value of $e$, we have to calculate the
sum of terms $[ S(Q_u(c))-S(Q_e(c)) ]$ 
for channels $c$ and $c'$. We have
\begin{eqnarray}
\lefteqn{ S(Q_u)-S(Q_e)+ S(Q_u')-S(Q_e') } \hspace{1cm}
\nonumber \\ 
&=& S(Q_e' Q^*)-S(Q_e)+ S(Q_e Q^*)-S(Q_e') \nonumber \\
&\le& 2 S(Q^*) \nonumber \\
&\le& 2 (n-2e)
\end{eqnarray}
where the last inequality reflects the limitation on $S(Q^*)$ imposed by
the dimension of the Hilbert space of $Q^*$. Thus each {\em pair} of terms
in the summation over $c$ can be bounded from above by $2(n-2e)$, so that 
a simple calculation yields an upper bound on the mutual entropy
of the joint ($n$-bit) channel
\begin{eqnarray}\label{eq_Ip5}
I(p) &\le& S(R) + \sum_{e=0}^{n} {n\choose e} (1-p)^{n-e} p^e (n-2e) 
\nonumber \\
&\le& k + n (1-2 p)
\end{eqnarray}
or a lower bound on the overall loss of the joint channel
\begin{equation}
L(p) \ge k + n (2p-1)
\end{equation}
This results in a lower bound on the {\it average} loss (per qubit)
for a quantum erasure channel of probability $p$ as a function of the
rate $R=k/n$,
\begin{equation}  \label{eq_lp}
l(p)= {L(p) \over n} \ge R + 2p -1
\end {equation}
(Note that this is only a lower bound on the loss, which is, by definition, a
non-negative quantity.) This inequality is consistent with
$l(p)+l(1-p)\ge 2R$ [cf. Eq.~(\ref{eq_Lp})].
It implies that a vanishing average loss (i.e., the
reliable transmission of information through the $n$-bit 
probabilistic channel) is only possible if the rate
\begin{equation}  \label{eq_QEC}
R \le 1-2p
\end{equation}
Thus, a quantum erasure channel with $p=1/2$ (i.e., if the channel is
erasing 50\% of the qubits) has a zero capacity.
Eq.~(\ref{eq_QEC}) confirms the linear interpolation\footnote{In 
Ref.~\cite{bib_bdsw}, it is shown that the capacity of a composite
channel (which is a convex combination of a perfect and an imperfect
quantum channel) cannot exceed the appropriately averaged capacity of these
two component channels. In other words, the quantum capacity cannot be
superadditive when ``mixing'' a perfect channel with an imperfect one.}
between the 50\%-erasure
channel (for which the capacity is zero) and a noiseless channel
(for which the capacity is one)
that was used in Refs.~\cite{bib_bdsw,bib_qec}. In addition, since it is shown 
in Refs. \cite{bib_bdsw,bib_qec} that this upper bound coincides with
a lower bound obtained from one-way random hash coding,
Eq.~(\ref{eq_QEC}) therefore describes
the {\em exact} capacity of the quantum erasure
channel $C=1-2p$~\cite{bib_qec}.

\subsection{Quantum depolarizing channel}

We now consider a quantum depolarizing channel with error probability $p$.
In this channel, each qubit interacts with the
environment such that it undergoes a bit-flip ($\sigma_x$ rotation),
a phase-flip ($\sigma_z$ rotation), or
the combination of both ($\sigma_y$ rotation) 
with probability $p/3$ each.
In Secs. IIIA and IIIB, we have seen that
the simple connection between quantum error- and erasure-correcting
codes provides a trivial relation between the resulting upper bounds
on the rate of channels subjected to a bounded fraction of errors or
erasures. Unfortunately, there is no such simple relationship
when comparing the quantum erasure and depolarizing channels. 
As a matter of fact, the upper bound on error-correcting codes, 
Eq.~(\ref{eq_rate_p_errors}), is not immediately applicable
to the quantum depolarizing channel.
Such a situation results from the fact that 
the definition of an error-correcting
code requires that {\it all} error patterns (of at most $t=pn$ errors) are
perfectly corrected, while a rate will be said to be attainable 
through a channel (i.e., it is
below the quantum channel capacity) whenever the fraction of uncorrected
error patterns is asymptotically vanishing (for $n\to \infty$).
Still, the reasoning used to calculate the Singleton bound on the
quantum erasure channel is
applicable to the quantum depolarizing channel as well.
\par

Assume that an individual qubit, $Q$, is initially entangled
with the reference, so that $RQ$ is in the state
$|\Psi_{RQ}\rangle$  (for example
the singlet state). After processing by the channel, the system
$RQ'$ is in a Werner state~\cite{bib_werner}
of ``entanglement fidelity'' $F=1-p$, that is, the mixed state:
\begin{equation}
\rho_{RQ'}=(1-4p/3) |\Psi_{RQ}\rangle\langle \Psi_{RQ}|
+ 4p/3 \left({{\bf 1}_R \over 2} \otimes {{\bf 1}_Q \over 2}\right)
\end{equation}
In other words, the qubit emerges at the output of the channel either in
a random state (having totally lost the entanglement with $R$) with
probability $4p/3$, or in its intact original state (fully entangled with $R$)
with probability $1-4p/3$. When $p=3/4$, the channel is 100\%
depolarizing, i.e., its output is random.
\par

In the joint $n$-bit channel, each qubit undergoes the above
evolution independently of the other ones. As before, the resulting
$n$-bit probabilistic channel can be described as a superposition
of binomial channels, in which each qubit is either kept unchanged
or ``randomized''. The distribution of the underlying channels
is thus a binomial one, just as in the previous Section,
the only difference being that $p$ is replaced here by $4p/3$.
The entire calculation of the previous Section can then be repeated,
because the ``randomization'' or ``erasure'' of a qubit are
equivalent as far at the mutual entropy with $R$ is concerned.
Indeed, for a channel $c$, we have
\begin{equation}
I_c=S(R{\rm:}Q_c')=S(R{\rm:}Q_u(c))
\end{equation}
where $Q_u(c)$ correspond to qubits that are not randomized (rather
than not erased) in channel $c$. This is obvious because the randomized
qubits [in state $(|0\rangle\langle 0|+|1\rangle \langle 1|)/2$]
are independent of $R$, just as the erased qubits (in state $|2\rangle$).
\par

First, in analogy with Eq.~(\ref{eq_Lp}), we have
\begin{equation}
L(p)+L(3/4-p) \ge 2k
\end{equation}
implying that $L(3/8)\ge k$, so that the quantum depolarizing channel
with probability $p=3/8$ cannot be lossless (in fact, it
has a vanishing capacity, as shown below). Equivalently, we
have $L(p) \ge I(3/4-p)$, showing that the $p=3/4$ channel cannot
transmit classical or quantum information, i.e., $I(3/4)=L(0)=0$.
The resulting Singleton lower bound on the {\it average} loss (per qubit) is 
\begin{equation}
l(p) = {L(p)\over n} \ge R + 8p/3 -1 
\end{equation}
in analogy with Eq.~(\ref{eq_lp}).
Consequently, the quantum depolarizing channel can have a
vanishing average loss (i.e., allow an asymptotically reliable
transmission of information by using blocks of $n$ qubits)
at the condition that
\begin{equation}  \label{eq_QDC}
R \le 1 - 8p/3
\end{equation}
as originally shown in Ref.~\cite{bib_bdsw}.
Thus the quantum depolarizing channel with $p=3/8$ has a zero capacity
for the transmission of quantum information. (Note that such a channel
corresponds in fact to a 50\%-depolarizing channel, where 50\% of the
qubits are replaced by a random qubit. This channel can obviously
not have a non-zero capacity, as a consequence of the no-cloning
theorem~\cite{bib_bdsw}.) As for the quantum erasure channel, a linear
interpolation between the perfect channel and the 50\%-depolarizing
channel can be used (and also results independently from the our reasoning).
Note that Eq.~(\ref{eq_QDC}) yields only an upper bound on the
capacity of the quantum depolarizing channel, which is provably
not achievable (in contrast with the equivalent bound for the quantum
erasure channel). Indeed, a tighter bound for the depolarizing channel
has been obtained very recently~\cite{bib_bruss},
\begin{equation}
R \le 1 - 4p    \label{eq_bruss}
\end{equation}
which is based on the Buzek-Hillery universal cloning
machine~\cite{bib_bh}. While Eq.~(\ref{eq_bruss}) happens
to be equivalent to Eq.~(\ref{eq_rate_p_errors}), 
there appears to be no direct relation between them.
A simple intuitive reason why this bound is stronger than
Eq.~(\ref{eq_QDC}) can be understood by realizing that
the two quantum channels underlying the universal cloning machine
(from the single input to both outputs) cannot be described
classically.\footnote{Remember that a standard no-cloning argument
to show that the rate vanishes at $p=3/8$ is based on a classical 
machine that is transmitting an input qubit to one of two outputs
with probability 1/2, a random qubit being sent on the other output,
which results in two 50\%-depolarizing channels~\cite{bib_bdsw}.
Clearly, this is the most constraining bound on $R$ that can be constructed
by use of such a ``classical cloning''.}
Indeed, when tracing over one of the outputs of the universal cloning
machine, the other output appears to emerge from a $p=1/4$ (or
$F=3/4$) channel, i.e., a 33\%-depolarizing channel. This
looks as if the qubit was sent with probability 2/3 to each
output of the cloning machine, which is obviously not understandable
in classical terms. Only a quantum superposition, involving the
cloning machine and both outputs, can account for this situation
and results in a stronger upper bound~\cite{bib_bruss}.
The information-theoretic analysis of quantum cloning (and related
entropic bounds)
will be the subject of further investigation.
\par

To our knowledge, the only stronger upper bound on the capacity
(for a restricted range of $p$ values) of the depolarizing channel
is $R\le 1-H(p)$, based on a connection between quantum additive
(or stabilizer) codes and classical linear codes~\cite{bib_cleve}.

\section{Quantum channel with an auxiliary classical channel.}

In this Section, we consider a quantum channel
which is supplemented with a classical channel (assumed to
be noiseless and of unlimited classical capacity). We are interested
in calculating the Singleton upper bound on the rate of reliable ($L=0$)
transmission of {\em quantum} information
through a noisy quantum channel, knowing that
a classical side channel can be used simultaneously for forward
communication only. In particular,
we aim at analyzing how our quantum information-theoretic formalism
accounts for the property that such a classical (one-way) 
communication channel does {\it not} increase the quantum capacity
of the noisy channel~\cite{bib_bdsw}.

\subsection{Entropic treatment of the channel}

\begin{figure}
\caption{Schematic representation of the ``encoder'' performing a
unitary transformation on the ``logical'' system $L$ (initially
entangled with $R$ with a mutual entropy $2k$). The outputs
are: the $n$ ``physical'' qubits $Q$, the classical bits $C$, and 
the ``precursor'' $P$ (``microscopic'' classical bits, before amplification).
Each classical bit can be thought of as a set of qubits that are
classically correlated when tracing over $P$, appearing then
as a collective classical variable (see \protect\cite{bib_measure}).}
\vskip 0.25cm
\centerline{\psfig{figure=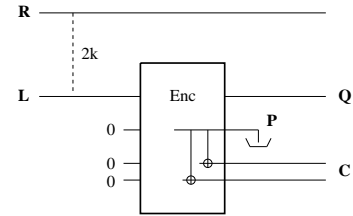,width=1.75in,angle=-90}}
\label{fig_encoder}
\vskip -0.25cm
\end{figure}

We first consider the problem in the language of quantum codes,
following closely Ref.~\cite{bib_cerfcleve}. As
explained earlier, the result will then be immediately applicable
to channels with an error model where the fraction of errors or erasures
is bounded. Also, the expression of a ``probabilistic'' joint channel as
a binomial superposition of underlying channels will then yield the
corresponding bounds for the quantum erasure or depolarizing channel.
We start with a ``logical'' system $L$ (logical words), which is
initially entangled with the reference $R$ (the initial mutual entropy
being $2k$, so that $k$ arbitrary qubits are transmitted). 
As before, the system $RL$ is initially in a pure entangled state
$|\Psi_{RL}\rangle$.
We assume that the encoding operation on $L$ includes a partial
measurement, so that the encoder has a quantum
and a classical output, as shown in Fig.~\ref{fig_encoder}.

\par

Note that the encoding process can be described in terms of
a unitary transformation applied on $L$ (supplemented
with an ancilla initially in a $|0\rangle$ state) without
any projection operators, so that the classical output
is simply described as a mixture of orthogonal
quantum states. More precisely, we have a quantum output $Q$
(consisting of $n$ ``physical'' qubits), $P$, which is the ``precursor''
of the classical variable, and $C$, which represents the
{\it amplified} classical variable. In general, 
$C$ is a discrete variable of arbitrary dimension, but we often
use the term classical bit (two-dimensional variable) in the following.
Each classical bit $C$ can be thought of as a set of qubits that become
{\it classically} correlated when tracing over the corresponding
``precursor'', and can then be viewed as a collective variable.
The resulting {\it amplification} gives to $C$ the appearance of
a classical variable (see \cite{bib_measure}).
It is convenient to keep the classical output
$C$ separate from its precursor $P$ for reasons that will appear later.
\par

To fix the ideas, let us write the intermediate wave function
of $RQP$ before amplification: 
\begin{equation}   \label{eq_psiRQP}
|\Psi_{RQP}\rangle = \sum_i \sqrt{p_i}\; |\psi_{RQ}^i\rangle \otimes
|\phi_P^i\rangle
\end{equation}
Here, the $|\phi_P^i\rangle$ form a set of orthogonal states
for $P$, and the $|\psi_{RQ}^i\rangle$ are orthogonal states of
$RQ$. Eq.~(\ref{eq_psiRQP}) is simply the Schmidt decomposition
of $|\Psi_{RQP}\rangle$ divided in $RQ$ versus $P$, and implies
$S(RQ)=S(P)$ before amplification. It is important to note that
the orthogonal states $|\phi_P^i\rangle$ for $P$ correspond to the
{\it classical} information that will be amplified (the
amplification of $P$ will be performed in this basis), the classical
variable being distributed according to $p_i$. 
Eq.~(\ref{eq_psiRQP}) is the most general expression of the output
of the encoder (before amplification of $P$) if we require that its overall
operation is unitary (the joint state of $RQP$ must be pure). We also
have obviously $S(R{\rm:}QP)=S(R{\rm:}L)=2k$ as a result of the
conservation of mutual entropy under a local unitary operation on both
subsystems~\cite{bib_physcomp}.
\par

The amplification of the precursor (symbolically represented by the
{\sc xor} gates inside the encoder) in the $|\phi_P^i\rangle$ basis
gives rise to a total wave function for the system $RQPC$ 
(at the output of the encoder) of the form:
\begin{equation}   \label{eq_psiRQPC}
|\Psi_{RQPC}\rangle = \sum_i \sqrt{p_i}\; |\psi_{RQ}^i\rangle \otimes
|\phi_P^i\rangle \otimes |\phi_C^i\rangle
\end{equation}
where the $|\phi^i\rangle$ correspond to an orthogonal set of states
for both $P$ and $C$. The classical information in $P$ has been
``amplified'', so that the precursor $P$ and the collective
set of qubits $C$ are fully
{\em classically} correlated when tracing over the remaining
variables. The entropy of the classical channel is $S(C)=S(P)=H[p_i]$.
In fact $P$ and $C$ are interchangeable, but
we need to keep them separate to account for the fact that 
amplifying the classical bits, i.e., tracing over the precursor
$P$, results in a mixed state for the systems $R$, $Q$, and $C$.
The density matrix for the system $RQC$ is given by
\begin{equation}
\rho_{RQC}=\sum_i p_i |\psi_{RQ}^i\rangle \langle\psi_{RQ}^i|
\otimes |\phi_C^i\rangle \langle \phi_C^i |
\end{equation}
This can be viewed as a {\it classical} mixture of orthogonal states
of $RQ$ and $C$. Thus, conditionally on the classical bits $C$, the
system $RQ$ is in a pure (generally entangled) state $|\psi_{RQ}^i\rangle$.
The question now will be, roughly speaking, to determine under which
circumstances the quantum output $Q$ (possibly altered by decoherence
or partially erased) retains the full entanglement with $R$ when it is
{\it supplemented} with the classical information $C$. In other words,
the question will be whether decoding {\it using} $C$ allows to 
perfectly recover the logical words
(and is more effective than in the absence of classical information).
\par

We define the two parameters $k$ and $c$ as
\begin{eqnarray}
S(R)&=&k              \label{eq_S(R)}\\
S(C)&=&S(P)=S(CP)=c   \label{eq_S(C)}
\end{eqnarray}
where the second equation is due to the fact that $C$ and $P$
are {\em fully} classically correlated, i.e, $S(C{\rm:}P)=c$. The parameter
$c=H[p_i]$ simply represents the Shannon entropy processed by
the classical side channel. As before, the parameter $k$
stands for the number of logical qubits (i.e., $k$-qubit arbitrary states
are encoded), or, equivalently, 
the source entropy when considering a quantum channel.
\par

\begin{figure}
\caption{Entropy diagram characterizing the reference $R$, the system $QC$
(quantum and classical output of the encoder), and $P$ (precursor
or the classical bits). The condition $S(R{\rm:}P)=0$ means that
the full mutual entropy $2k$ with $R$ is found in $QC$ and does
not leak out when tracing over $P$, so that the amplification
is {\em lossless}. The two parameters are $S(R)=k$ and $S(P)=c$.}
\vskip 0.25cm
\centerline{\psfig{figure=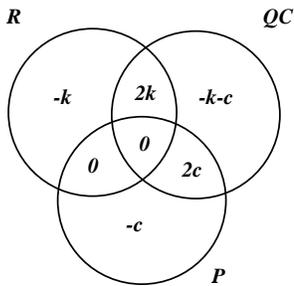,width=1.50in,angle=-90}}
\label{fig_QC}
\vskip -0.25cm
\end{figure}

A necessary condition for the global (classical + quantum) channel
to be lossless is clearly that the amplification of $P$ is a
{\em lossless} process, that is, does not destroy the quantum coherence
of the logical words. More precisely, the
constraint we must express is that the channel $L \to QC$
(considered as a quantum channel) is {\em lossless}, that is
\begin{equation}  \label{eq_leak}
S(R{\rm:}P)=0
\end{equation}
where $P$ plays the role of an ``environment'' for this channel.
This means that, when amplifying the classical bits $C$
(ignoring the precursor $P$)
no entanglement with $R$ is lost. In other words, the
mutual entropy $2k$ with $R$ is entirely found in $QC$, 
i.e., $S(R{\rm:}QC)=2k$, so that no entanglement with $R$ leaks out
in $P$ when tracing over $P$.
(This is so even though the joint system $QPC$ is fully entangled
with $R$, i.e., $S(R{\rm:}QPC)=2k$.) 
The condition (\ref{eq_leak}) implies that
\begin{eqnarray}
S(QC)&=&S(RP) \nonumber \\
     &=&S(R)+S(P)-S(R{\rm:}P) \nonumber \\
     &=&k+c
\end{eqnarray}
Thus, the parameters $k$ and $c$ fully determine the ternary
entropy diagram\footnote{The ternary diagram of a tripartite system 
in a pure state is determined in general by three parameters, for example
the reduced entropy of each of the three
components~\cite{bib_physcomp}.} for $R$, $QC$, and $P$,
as shown in Fig.~\ref{fig_QC}.
Note that $C$ and $P$ are interchangeable, so that we have
also $S(QP)=k+c$.
\par

\begin{figure}
\caption{Entropy diagram characterizing the reference $R$, the
quantum system $Q$ (output of the encoder), and $PC$
(the classical output and precursor {\em before} amplification).
The three parameters are $S(R)=k$, $S(Q)=s$, and $S(CP)=c$.}
\vskip 0.25cm
\centerline{\psfig{figure=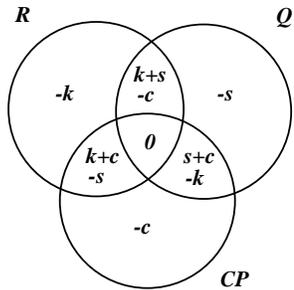,width=1.50in,angle=-90}}
\label{fig_CP}
\vskip -0.25cm
\end{figure}

A fourth parameter is necessary to fully describe the entropies
of the 4-partite system ($R$, $Q$, $C$, and $P$):
\begin{equation}
S(Q)=s   \label{eq_S(Q)}
\end{equation}
It corresponds to the von Neumann entropy of the quantum input
of the noisy channel (or quantum output of the encoder).
Grouping $C$ and $P$, we can also display the ternary entropy
diagram for $R$, $Q$, and $CP$, using the fact that the joint system
is in a pure state (cf. Fig.~\ref{fig_CP}). It shows
that neither the quantum output $Q$, nor the classical one $CP$ {\em before}
amplification (i.e., including
the precursor) is unentangled with $R$. In short, the situation
is that $P$ alone is unentangled with $R$ (it can be traced over without
altering the entanglement with $R$), while $P$ and $C$ together are
entangled with $R$. Moreover, even though the classical output
can be amplified without losing coherence, the classical information $C$
is in general necessary (together with $Q$) in order to recover
the initial entanglement of $L$ with respect to $R$,
as implied by $S(R{\rm:}QC)=2k$.
\par

\begin{figure}
\caption{Entropy diagram characterizing the reference $R$, the
quantum system $Q$ (output of the encoder), and the classical output 
$C$ {\em after} amplification. The joint system is in a {\em mixed} state
of entropy $S(RQC)=c$. Note that $R$ is independent of $C$ (i.e., $C$
contains no information about the encoded logical word), while
$RQ$ is (fully) classically correlated with $C$.}
\vskip 0.25cm
\centerline{\psfig{figure=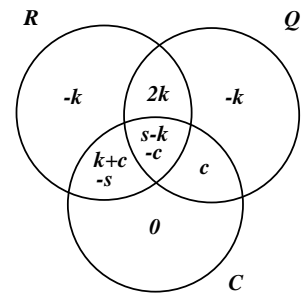,width=1.50in,angle=-90}}
\label{fig_RQC}
\vskip -0.25cm
\end{figure}

We now want to describe the entropic situation {\em after} amplification
of the classical bits, i.e., after tracing over the precursor $P$.
The density matrix for the system $RQC$ (or equivalently for the
system $RQP$) is given by
\begin{equation}
\rho_{RQC}=\sum_i p_i |\psi_{RQ}^i\rangle \langle\psi_{RQ}^i|
\otimes |\phi_C^i\rangle \langle \phi_C^i |
\end{equation}
The system $RQ$ is thus {\it classically} correlated with $C$, and
we have
\begin{equation}  \label{eq_RQ:C}
S(RQ{\rm:}C)=S(RQ)=S(C)=c
\end{equation}
For each value $i$ of the classical bits
(occurring with probability $p_i$), $RQ$ is in a given pure 
(generally entangled) state $|\psi_{RQ}^i\rangle$.
Now, using $S(QP)=S(QC)=k+c$,
we can show that $R$ is {\em independent} of $C$:
\begin{eqnarray}  \label{eq_R:C}
S(R{\rm:}C)&=&S(R)+S(C)-S(RC)  \nonumber \\
           &=&S(R)+S(C)-S(QP)  \nonumber \\
           &=& k + c - (k+c) =0
\end{eqnarray}
This means that, if the amplification
of the classical bits is {\it lossless} (that is,
cannot result in an irrecoverable lost
of mutual entanglement with $R$), then the ``amplified'' classical bits $C$
must be statistically {\it independent} of $R$. In other words, no
information about the encoded logical word is found in the
``amplified'' classical bits $C$ even though $C$ is in general necessary
to recover the entanglement with $R$.
Using $S(QC)=S(R)+S(C)$, the mutual entropy between
$Q$ and $C$ can be written as
\begin{eqnarray}  \label{eq_Q:C}
S(Q{\rm:}C)&=&S(Q)+S(C)-S(QC)  \nonumber \\
           &=&S(Q)-S(R)        \nonumber \\
           &=&s-k
\end{eqnarray}
Thus, the quantum output $Q$ is in general {\it not} independent
of the classical bits $C$ (in contrast with $R$). This simply means that,
in general, the encoder can introduce some extra entanglement 
between $Q$ and $CP$, additionally to the initial entanglement $2k$ between
$Q$ and $R$, giving rise to a non-vanishing
mutual entropy. However, this additional entanglement is useless as far as
the transmission of quantum information is concerned, and we will see
below that the interesting situation corresponds to $s=k$, in which case
$S(Q{\rm:}C)=0$.
Finally, it is easy to see that $RQC$ is
in a mixed state of entropy $S(RQC)=S(P)=c$. The latter condition,
together with
Eqs.~(\ref{eq_S(R)}), (\ref{eq_S(C)}), (\ref{eq_S(Q)}), 
(\ref{eq_RQ:C}), (\ref{eq_R:C}), and (\ref{eq_Q:C}),
fully describes the entropies of the tripartite
system $RQC$ after tracing over $P$ (i.e., after
amplification). The corresponding ternary entropy diagram
is presented in Fig.~\ref{fig_RQC}.

\par

In summary, we have $S(R)=k$, $S(Q)=s$, $S(C)=c$, $S(RQ)=c$,
$S(RC)=k+c$, $S(QC)=k+c$, and $S(RQC)=c$, which implies that
\begin{eqnarray}
S(R{\rm:}Q{\rm:}C)&=&S(R)+S(Q)+S(C)-S(RQ) \nonumber\\
& &-S(RC)-S(QC)+S(RQC)\nonumber\\
&=& s-k-c
\end{eqnarray}
As visible from this diagram, the system $RQ$ is in a pure entangled
state {\em conditionally} on the amplified classical variable $C$, 
i.e., $S(RQ|C)=0$, with a characteristic
diagram $(-k,2k,-k)$. Also, $RQ$ is classically correlated with $C$,
with a diagram $(0,c,0)$,
so that the simultaneous knowledge of $R$ and $Q$ yields $C$,
i.e., $S(C|RQ)=0$. Finally, it is easy to check that the diagram
for $R$ vs. $C$ is $(k,0,c)$, i.e., $C$ is independent of $R$.
\par

Several inequalities relating the parameters $k$, $c$, and
$s$, must be satisfied. First, the subadditivity of entropies implies that
\begin{eqnarray}
S(R{\rm:}Q)&=&k+s-c \ge 0 \\
S(Q{\rm:}C)&=&s-k \ge 0
\end{eqnarray}
Similarly, the strong subadditivity of entropies implies that
\begin{equation}
S(R{\rm:}C|Q)=k+c-s \ge 0
\end{equation}
These inequalities can be summarized as
\begin{equation}
0 \le s-k \le c \le s+k
\end{equation}
implying namely that $0\le c \le 2s$.
The two limiting cases are (i) $c=0$ and $s=k$, which corresponds to
a quantum channel without a classical side channel; and (ii)
$c=2s=2k$, which corresponds to the situation where the entropy
of the classical channel is maximum. In the following, we will focus on
case (ii), that is, when a maximum amount of classical information
is processed,
since it is supposedly the case where the classical side channel
might help the transmission of quantum information the most.

\begin{figure}
\caption{Entropy diagrams {\em before} amplification of the classical bits
in the case where the processed classical information is maximum
($c=2s=2k$). As before, $S(R{\rm:}P)=0$, so that
the amplification of the precursor $P$ does not affect quantum coherence.
Note that $S(R{\rm:}Q)=0$, so that the the quantum output
$Q$ can be erased if no amplification is performed.}
\vskip 0.25cm
\centerline{\psfig{figure=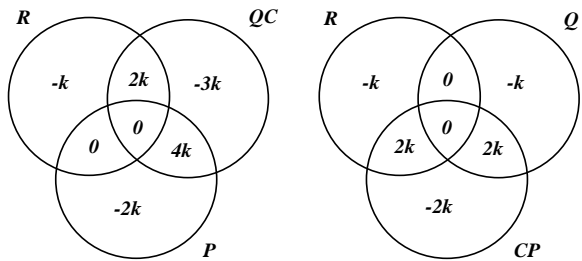,width=3.0in,angle=-90}}
\label{fig-2k}
\vskip -0.25cm
\end{figure}

\begin{figure}
\caption{Entropy diagram {\em after} amplification of the classical bits
in the case where the processed classical information is
maximum ($c=2s=2k$).
Here, $R$ and $Q$ play the same role. The classical variable
$C$ contains the information about which entangled state $RQ$ is in,
while $R$ or $Q$ alone is independent of $C$.
This diagram with $k=1$ plays a crucial role in the entropic description of
superdense coding and teleportation, 
as shown elsewhere\protect\cite{bib_ca_teleport}.}
\vskip 0.25cm
\centerline{\psfig{figure=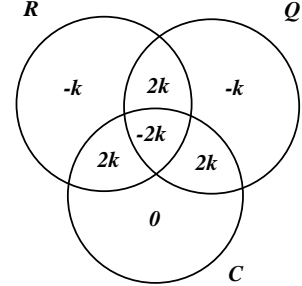,width=1.50in,angle=-90}}
\label{fig-teleport}
\vskip -0.25cm
\end{figure}

We display in Figs.~\ref{fig-2k} and \ref{fig-teleport}
the entropy diagrams corresponding to the limiting case $c=2k=2s$.
Note that $S(R{\rm:}Q)=0$ as shown in Fig.~\ref{fig-2k}, so that
the channel $L\to CP$ is lossless. Thus, the entire entanglement with $R$
is retained in the {\em unamplified} classical variable $CP$.
We will discuss this
below. Fig.~\ref{fig-teleport} implies that
$S(R{\rm:}C)=S(Q{\rm:}C)=0$, i.e., $Q$ and $R$ are both {\it independent}
of the classical variable $C$. This emphasizes the fact that
$R$ and $Q$ play exactly the same role in this limiting case (ii).
The peculiar feature here is that $Q$ and $R$ together are fully correlated
with $C$, according to the diagram $(0,2k,0)$,
although $Q$ or $R$ taken separately is independent of $C$. In other words,
the classical variable $C$ contains the information about which
entangled state $RQ$ is in (i.e., the mutual entropy $2k$) while
it contains no information about $Q$ or $R$ alone.
\par

The diagram in Fig.~\ref{fig-teleport} plays an important role in the
information-theoretic description of
quantum teleportation and superdense coding, as shown in a further
work~\cite{bib_ca_teleport}. In the special case where $k=1$
(no block coding is used), it describes quantum teleportation
in the following sense:
$Q$ is the particle that is initially sent to Bob
(one half of the Bell state shared with Alice),
and $C$ are the 2 classical bits 
that Alice sends to Bob. If the teleported particle (here $L$) 
is initially entangled with $R$ so that $RL$ is in a Bell state,
then $Q$ ends up in an entangled state 
with $R$ which is one of the four Bell states, {\em conditionally} on $C$. 
(Teleportation is completed by having Bob applying a unitary
transformation on $Q$ that is specified by $C$.)
Thus, the 2 classical bits code for one of the four Bell states, and the
corresponding diagram is shown in Fig.~\ref{fig-teleport} with $k=1$.
The amplification of the classical bits by Alice does {\it not}
destroy coherence, since we have $S(R{\rm:}P)=0$, and bringing 
$Q$ and $C$ together yields the initial entanglement with respect
to $R$, i.e., $S(R{\rm:}QC)=2$. The non-classical feature here
is that the latter equation can be satisfied even though
$S(R{\rm:}Q)=0$ holds at the same time, that is, the particle that
Bob received initially is independent of $R$ (this must be true as
a consequence of causality). Thus,
the entire entanglement with $R$ is carried by $C$, as reflected by
$S(R{\rm:}C|Q)=2$. This will be discussed
in more details elsewhere~\cite{bib_ca_teleport}.
\par

Note finally that the vanishing mutual entropy between $R$ and $Q$
implies that no entanglement with $R$ is found in $Q$ alone.
Therefore, the quantum output $Q$ can be erased without loosing
the entanglement with $R$, provided that the classical variable is
not amplified (by keeping $CP$).
In other words, since $S(R{\rm:}Q)=0$,
the knowledge of the classical bits $C$ (along with the precursor $P$)
is sufficient to recover $L$, even in the absence of $Q$. For example, in
teleportation, the {\it unamplified} classical bits alone are enough
to teleport an arbitrary state, so that the qubit $Q$ that Bob
received initially can be
erased. (Of course, this is unrealistic, since one never access to all the
microscopic degrees of freedom making the classical bits. Tracing over
one of them is enough to loose the quantum information if
$Q$ is erased.)

\subsection {Singleton bound on a quantum channel with a classical
side channel}

Let us now repeat the reasoning which results in the Singleton
bound on a quantum code (Section III), but taking into account the classical
auxiliary channel. Thus, we assume 
again that the quantum output $Q$ of $n$ qubits
is partitioned into an erased piece $Q_e$ (of $e$ qubits) and an unerased
one $Q_u$ (of $n-e$ qubits). We are seeking for a necessary condition
for the possibility of recovering the erasure of $Q_e$
when the decoder has access to the classical information $C$
(so the decoding operation can be conditional on $C$). As before, we consider
two different partitions of $Q$ (see Fig.~\ref{fig_cerfcleve}), and
express a lower bound on the entropy of the ``overlap'' $Q^*$.
The basic constraints (which must be satisfied simultaneously) are
\begin{eqnarray}
S(R{\rm:}Q_eP)  &=& 0  \\
S(R{\rm:}Q_e'P) &=& 0
\end{eqnarray}
expressing the fact the no entanglement (with respect to $R$) is
lost when amplifying the classical bits {\it and} erasing $Q_e$ (or $Q_e'$).
Equivalently, the full initial entanglement of $L$ must be
``squeezed'' into $Q_u$ (or $Q_u'$) and $C$:
\begin{eqnarray}
S(R{\rm:}Q_uC)  &=& S(R{\rm:}QC) = 2 S(R) \label{eq_R:Q_uC}\\
S(R{\rm:}Q_u'C) &=& S(R{\rm:}QC) = 2 S(R) \label{eq_R:Q_uC2}
\end{eqnarray}
In other words, the knowledge of the unerased part $Q_u$ (or $Q_u'$)
is sufficient to reconstruct (using $C$) the initial logical word.
Since the system $RQ_uQ_ePC$ is in a pure state, we have
\begin{equation}
S(RQ_uC)=S(Q_eP)=S(Q_eC)
\end{equation}
where we used the fact that $P$ and $C$ are interchangeable.
Now, dividing $Q_u$ into $Q_e'$ and $Q^*$, we can write an upper bound
on the mutual entropy between $R$ and $Q_uC$,
\begin{eqnarray}  \label{eq_R:Q_uC3}
S(R{\rm:}Q_uC)&=&S(R)+S(Q_uC)-S(RQ_uC)   \nonumber \\
              &=&S(R)+S(Q_e'Q^*C)-S(Q_eC)   \nonumber \\
              &\le& S(R)+S(Q_e'C)+S(Q^*)-S(Q_eC)
\end{eqnarray}
where we have used the subadditivity of quantum entropies.
Eqs.~(\ref{eq_R:Q_uC}), (\ref{eq_R:Q_uC2}), (\ref{eq_R:Q_uC3}),
and its counterpart (when $Q_u$ is replaced by $Q_u'$) thus give
\begin{eqnarray}
S(R)-S(Q^*) &\le& S(Q_e'C)-S(Q_eC) \\
S(R)-S(Q^*) &\le& S(Q_eC)-S(Q_e'C)
\end{eqnarray}
Combining these two last inequalities results in the same
inequality as in the case where no classical auxiliary channel
is used:
\begin{equation}
S(R) \le S(Q^*) \le n -2e
\end{equation}
Therefore, we obtain the {\it same} quantum Singleton bound
for a quantum code supplemented with a noiseless classical channel
as in the absence of such a channel:
\begin{equation}
k \le n-2e
\end{equation}
In other words, the classical side channel (transmitting data with an
entropy up to twice the quantum source entropy $k$)
does not increase the Singleton bound on the
maximum attainable distance for quantum codes.
\par

This result can be immediately applied to a quantum channel characterized
by a $p$-bounded fraction of erasures (cf. Section IIIA)
and supplemented with an
auxiliary classical channel, since, in that case, the use of a
$((n,k))$ code protecting for $e=np$ erasures is enough to guarantee
reliable transmission. Therefore, the upper bound on the rate is
given by
\begin{equation}
R \le 1 - 2p 
\end{equation}
similar to Eq.~(\ref{eq_rate_p_erasures}),
confirming the fact that the classical side channel does not
enhance the quantum information transmission through the quantum
channel~\cite{bib_bdsw}.
\par

In the case of a quantum erasure channel (with erasure probability
$p$) supplemented with a classical channel, the entire
reasoning of Section~IIIC can be repeated, the only difference being
that one has to calculate the sum of $[S(Q_uC)-S(Q_eC)]$ for 
two overlapping channels ($c$ and $c'$):
\begin{eqnarray}
\lefteqn{ S(Q_uC)-S(Q_eC)+ S(Q_u'C)-S(Q_e'C) } \hspace{1cm}
\nonumber \\ 
&\le& 2 S(Q^*) \nonumber \\
&\le& 2 (n-2e)
\end{eqnarray}
The resulting bound on the mutual entropy is thus
the same as Eq.~(\ref{eq_Ip5}), so that we have the same upper bound
on the rate of reliable transmission of quantum information:
\begin{equation}
R \le 1 -2p
\end{equation}
Finally, the cases of a channel with a $p$-bounded fraction of errors
and the quantum depolarizing channel can be treated exactly as
in Sections~IIIB and IIID, so that the classical side channel does
not modify the Singleton upper bound on the rate in both cases.

\section{Conclusion}

The search for better bounds on the capacity
of quantum channels such as the depolarizing channel
is still a major endeavor in quantum information theory
today. Clearly, entropic considerations alone do not suffice
to prove that a reliable quantum coding scheme exists with achieves
a transmission rate arbitrarily close to the capacity. As a matter of fact,
a similar situation prevails for classical channels as well.
Nevertheless, an entropic approach appears to be helpful in order to derive
bounds on the capacity of
classical or quantum channels from similar principles,
and to analyze classical and quantum communication in a unified
framework, as shown in this work. More generally,
the leading idea underlying the approach to quantum information
presented in this work and in 
Refs.~\cite{bib_channel,bib_neginfo,bib_oviedo,bib_physcomp,bib_accessible}
is to build a theory that extends Shannon's concepts to
the quantum regime.
Rather than attempting to define a {\em distinct} (purely quantum)
information theory that would apply to the transmission of quantum 
states only, we prefer
to consider an {\em extended} Shannon theory, which should account
for the processing of classical as well as quantum information
(arbitrary classical or quantum states). After all,
any classical information-processing system should be describable,
in principle, in terms of its underlying quantum mechanical degrees 
of freedom. In this sense, Shannon theory should simply be viewed 
as a special case of a more general theory of information in quantum
mechanics that remains to be built.
\par

The central characteristics of a quantum extension of Shannon theory
happens to be that {\em negative} von Neumann conditional entropies
must be used in order to encompass quantum entanglement in an 
information-theoretic description. It is pointed out in Ref.~\cite{bib_neginfo}
that this apparently innocent observation should be viewed as
the central novelty of a quantum mechanical
extension of Shannon theory beyond its original range.
Since most of the classical concepts of Shannon theory 
have a straightforward quantum analog, it is possible
to repeat a great part of the classical reasoning and apply it to
quantum information processes, as shown in this paper and
our previous work. More specifically, we have shown here
that an information-theoretic description of 
noisy quantum channels following closely
Shannon theory provides insight into the derivation
of entropic bounds on the quantum capacity (the maximum rate at
which quantum information can be reliably processed in spite
of the noise). Namely, the entropic Singleton bound on quantum
error-correcting codes~\cite{bib_cerfcleve} can be used
in order to investigate standard quantum channels such
as the quantum erasure channel or the quantum depolarizing channel.
The same formalism can be extended in order to account for an auxiliary 
classical channel used for forward communication besides the noisy
quantum channel. Entropic Singleton bounds can be derived 
in the latter case too, showing that the classical channel does not enhance
the quantum capacity (or an upper bound on it), in agreement 
with what was proven in Ref.~\cite{bib_bdsw}.
\par

The central part of the reasoning consists in calculating
a lower bound on the {\em average loss} of the channel (i.e., the loss
of the joint channel made of $n$ consecutive uses of
a memoryless channel, divided by the number of processed symbols $n\to\infty$)
which characterizes the ``quality'' of the transmission. If 
the use of block coding makes the joint ($n$-bit) channel lossless 
(i.e., the average loss
is zero), then reliable transmission of quantum information is achievable.
This is true even though the {\em one-symbol
loss} (for a single use of the channel) is non-zero, reflecting
the alteration due to noise in each use of the channel.
Perfect transmission by block coding is thus possible provided that
this lower bound on the average loss is zero or less, which results
in an upper bound on the attainable rate.
\par

Obviously, there remains much to be done in order to derive
better bounds on the rate (or perhaps the exact capacity)
using such an entropic approach.
We have made a progress in this direction, as illustrated by
the entropic Singleton bound on the capacity of the quantum erasure
channel ($C\le 1-2p$) which happens to be the exact capacity calculated
in~\cite{bib_qec}. For the quantum depolarizing channel, however,
we obtain a well-known bound on the capacity ($C\le 1-8p/3$, 
see~\cite{bib_bdsw}) which has been recently shown not to be
attainable~\cite{bib_bruss}. Nevertheless, the characterization of the
exact quantum capacity of the depolarizing channel is still an open problem,
and it is possible that the entropic approach presented here
could be further improved.
Also, the issue of the attainable capacity of a general noisy quantum
channel might be explored along the same lines. The search for
better entropic bounds on the capacity of quantum channels
will be the subject of future work.

\acknowledgements
We acknowledge C. Adami for numerous useful discussions.
This research was supported in part by the National Science Foundation
under Grant Nos. PHY 94-12818 and PHY 94-20470,
and by a grant from DARPA/ARO through the QUIC Program
(\#DAAH04-96-1-3086). 

\appendix
\section{Information-theoretic characterization
of a noisy classical channel}
In this Appendix, we outline the information-theoretic description
of noisy classical channels,
for the sake of clarifying the correspondence with the treatment
of noisy quantum channels used throughout this paper.
At first sight, a classical channel seems very different from a quantum
channel as no classical ``reference'' is used to purify
the input. Also, the classical input-output joint 
probability distribution has no quantum
equivalent, since there is no joint state for the initial quantum system
$Q$ and the final system $Q'$ (it is the {\em same} system).
However, a classical channel can be thought of as a device which
processes classical {\em correlation} (with respect to a reference R). If the
input $X$ is initially fully correlated with a reference $R$, then 
the residual mutual entropy between the output $Y$ and $R$ is a measure
of the amount of correlation (or information) transmitted through the channel. 
For a quantum channel, we
consider the processing of {\em entanglement} (with respect to $R$)
rather than correlation,
so that the residual mutual entropy between the decohered quantum
system $Q'$ (the quantum output) and $R$ is the interesting quantity.
This makes the classical-quantum correspondence easier to understand.
\par

A noisy classical channel with input $X$ and output $Y$ is characterized by
\begin{eqnarray}
I &=& H(X{\rm:}Y)  \\
L &=& H(X|Y)  \\
N &=& H(Y|X)
\end{eqnarray}
where $I$, $L$, and $N$ denote the {\em information}, the {\em loss}, and the
{\em noise}, respectively (see, e.g.,~\cite{bib_ash}). Information
processing through the channel is measured by the balance
between $I$ and $L$, these two quantities summing to the source
entropy:
\begin{equation}
I+L = H(X)
\end{equation}
The loss measures the inherent uncertainty in the process of inferring
the input of the channel from the altered output (decoding), that is,
the entropy of the input conditional on the output.
When the loss $L$ is zero ({\em lossless} channel), the information
$I$ is maximum so that classical
information is perfectly transmitted through the channel. Conversely,
when $I=0$ (and $L$ is maximum), no classical information is processed
by the channel.
The noise $N$ reflects the uncertainty of the output symbol for a given
input symbol, and is irrelevant as far as the transmission of
information is concerned (it corresponds to the loss of the {\em reverse}
channel where the input and output are interchanged).
A channel with $N=0$ is called {\em deterministic},
and a channel which is both deterministic and lossless is named
{\em noiseless}. We follow the same nomenclature for quantum channels in
this paper.

If $n$ {\em independent} channels are used in parallel with $X_1\cdots X_n$
as an input string and $Y_1 \cdots Y_n$ as an output string, it can
be shown that the information and the loss are {\em subadditive}:
\begin{eqnarray}
I &\le& I_1 + \cdots + I_n  \label{eq_subadI} \\
L &\le& L_1 + \cdots + L_n  \label{eq_subadL}
\end{eqnarray}
where $I$ ($L$) is the information (loss) of the joint ($n$-symbol)
channel, while $I_i$ ($L_i$) is the information (loss) of the
$i$-th individual channel $X_i \to Y_i$. 
Property (\ref{eq_subadI}) results from the subadditivity of Shannon entropies
\begin{equation}
H(Y_1\cdots Y_n) \le H(Y_1)+\cdots +H(Y_n)
\end{equation}
and from the fact that the channels are independent (each $Y_i$
depends on $X_i$ only)~\cite{bib_cover}:
\begin{eqnarray}
I&=& H(X_1\cdots X_n {\rm:} Y_1 \cdots Y_n)   \nonumber \\
 &=& H(Y_1\cdots Y_n) - H(Y_1 \cdots Y_n|X_1\cdots X_n)  \nonumber \\
 &\le& \sum_{i=1}^n \Big[ H(Y_i) 
        - H(Y_i | Y_1\cdots Y_{i-1},X_1\cdots X_n) \Big] \nonumber \\
 &\le& \sum_{i=1}^n \Big[ H(Y_i)- H(Y_i|X_i) \big] 
\end{eqnarray} 
Property (\ref{eq_subadL}) is an immediate consequence of
the subadditivity of Shannon conditional entropies
\begin{equation}
H(X_1\cdots X_n|Y_1 \cdots Y_n) \le H(X_1|Y_1)+\cdots +H(X_n|Y_n)
\end{equation}
Since the information and the loss of each individual (one-symbol)
channel sum to the source entropy of that channel
\begin{equation}
I_i+L_i = H(X_i)
\end{equation}
the allowed range for the overall loss of the
joint channel is
\begin{equation}
L_1+\cdots +L_n - M \le L \le L_1+\cdots +L_n
\end{equation}
with $M=H(X_1)+\cdots +H(X_n)-H(X_1 \cdots X_n) \ge 0$.
Consequently, the loss cannot increase by using block coding
(i.e., using parallel channels with correlated input symbols),
but it can decrease by an amount which is bounded by $M$.
Note that $M$ vanishes when the input symbols are independent,
while a positive value of $M$ reflects the correlation between
the input symbols.
\par

It is simple to obtain a necessary condition for perfect
transmission (i.e., with a vanishing overall loss)
by block coding through a noisy channel. Clearly, the condition
\begin{equation}
H(X_1\cdots X_n) \le H(X_1)+\cdots +H(X_n)-L_1-\cdots -L_n
\end{equation}
must be fulfilled for the lower bound on $L$ to extend to zero.
Therefore, the rate of transmission through the joint channel,
$R=H(X_1\cdots X_n)/n$ is bounded from above by the averaged
{\em one-symbol} information of the individual channels:
\begin{equation}
R \le {I_1 + \cdots + I_n \over n}
\end{equation}
This is related to the weak converse of Shannon's noisy coding 
theorem~\cite{bib_cover}:
the transmission cannot be perfect (or lossless, i.e., $L=0$)
if the rate of transmission exceeds the (averaged) mutual Shannon
entropy characterizing each use of the channel. This is consistent
with Shannon's result that the classical channel capacity is the
maximum (over all input distributions) of the mutual information
between channel input and output for a {\it single} use of the channel.
Note that
entropy considerations alone do not suffice to prove that a reliable
coding scheme exists with achieves a transmission rate arbitrarily
close to the capacity. A similar situation is found
for quantum channels as well, as shown throughout this paper.
Still, an entropic approach is very helpful in order to derive
bounds on classical or quantum channels from basic principles,
and to analyze classical and quantum communication in a unified manner.

\end{multicols}
\end{document}